\documentclass[smallextended]{svjour3}

\smartqed

 \usepackage{graphicx}
 \usepackage{fix-cm}
 \usepackage{epsfig}
 \usepackage{epstopdf}
 \usepackage{graphics, amsmath, amssymb}
\usepackage{graphicx,subfigure,color}
\usepackage{soul}
\usepackage{amssymb}
\usepackage{amssymb}
\usepackage{amsmath}
\usepackage{mathrsfs}
\usepackage{caption,setspace}
\usepackage{braket}
\usepackage{amsmath,amssymb,graphicx}
\usepackage{float}
\usepackage{graphicx}
\usepackage{mathtools}

\journalname{Quantum Information Processing}
\begin{document}

\title{Dissipative quantum repeater}

\titlerunning{Dissipative quantum repeater}

\author{M. Ghasemi \and M. K. Tavassoly}

\institute{M. Ghasemi \at
                aff 1  Atomic and Molecular Group, Faculty of Physics, Yazd University, Yazd  89195-741, Iran\\
                         \email{m.ghasemi@stu.yazd.ac.ir}
 %==========================================================================================
           \and
 %==========================================================================================
           M. K. Tavassoly \at
              aff 1  Atomic and Molecular Group, Faculty of Physics, Yazd University, Yazd  89195-741, Iran\\
              Tel.:  +98 35 31232774 \\
              Fax:  +98 35 31230132 \\
              \email{mktavassoly@yazd.ac.ir}}
%==========================================================================================
 %          \and
 %==========================================================================================
 \date{Received: date / Accepted: date}

 \maketitle

 \begin{abstract}
  By implementing a quantum repeater protocol, our aim in this paper is the production of entanglement between two two-level atoms locating far from each other.
     To make  our model close to experimental realizations, the atomic and field sources of dissipations are also taken into account.
     We consider eight of such atoms (1, 2, ..., 8) sequentially located in a line which begins (ends) with atom $1$ ($8$).
     We suppose that, initially the four atomic pairs $(i,i+1)$, $i=1,3,5,7$ are mutually prepared in maximally entangled states.
    Clearly,  the atoms $1, 8$, the furthest atoms which we want to entangle them  are never entangled, initially.
        To achieve the purpose of paper, at first we perform the interaction between the atoms $(2,3)$ as well as $(6,7)$ which results
       in the entanglement creation between $(1,4)$ and $(5,8)$, separately. In the mentioned interactions we take into account
        spontaneous emission rate ($\Gamma$) for atoms and field decay rate from the cavities ($\kappa$)  as two important and unavoidable dissipation sources.
         In the continuation, we transfer the entanglement to the objective pair $(1,8)$ by two methods: \textit{i}) Bell state measurement (BSM), and \textit{ii})
          cavity quantum electrodynamics (QED). The successfulness of our protocol is shown via the evaluation of concurrence as the well-established measure
           of entanglement between the two (far apart) qubits $(1,8)$. We also observe that, if one chooses the cavity and the atom such that $\kappa=\Gamma$ holds,
            the effect of dissipations is effectively removed from the entanglement dynamics in our model.
            In this condition, the time evolutions of concurrence and success probability are regularly periodic. Also, concurrence and success probability reach to their maximum values in a large time interval by decreasing the detuning in the presence of dissipation.
 \keywords{Quantum repeater \and Entanglement swapping \and Atom-field interaction \and Dissipation source.}
 % %\PACS{03.65.Yz \and 03.67.Mn \and 42.50.-p}
 \end{abstract}
 \section{Introduction}
 \label{intro}

The idea of quantum repeater, firstly proposed by Briegel et al. \cite{Briegel1998} in 1998 and then widely spread between the quantum information science and technologies researches \cite{Dur1999,Duan2001,Jiang2014}. That is a method for the distribution of entanglement and entangled states over large distances, \textit{i.e.}, creating entanglement between  non-entangled separated atoms, that these transferred entangled states are useful in quantum key distribution \cite{Curty2004}, quantum cryptography \cite{Jennewein2000} and quantum communication \cite{Hsieh2003,Shi2011}.
In recent decades significant efforts have been paid to the implementation of a practical quantum repeater. For instance, the entanglement has been established between two fixed single-atom quantum memories separated by 1 meter \cite{Moehring2007}. Also, the entanglement has been achieved between solid-state qubits separated by 3 metres \cite{Bernien2013}, and the authors in \cite{Hofmann2012} reported that the entanglement between spins of two single Rubidium-87 atoms trapped independently 20 meters apart has been established.
In quantum repeater protocols the path of process is generally divided into a number of shorter parts and the entanglement is transferred by entanglement swapping processes \cite{Eghbali2017,Zukowski1993,Ghasemi2016,Ghasemi2017,Nourmandipour2016,nourmandipour2015} from a pair of original particle state to a far destination particle pair state. Bell state measurement (BSM) \cite{Liao2011} and cavity quantum electrodynamics (QED) \cite{Haroche1999,Chun2006} are two well-known ways for swapping the entanglement. In the BSM method, the entanglement is swapped  from an entangled state to non-entangled particles. However, in the QED method, entanglement swapping is achieved by performing an interaction between two non-entangled separate particles \cite{Yang2005,Pakniat2016,Pakniat2017}, while these interactions are governed by (generalized) Jaynes-Cummings  (JCM) \cite{Jaynes1963} or sometimes Tavis-Cummings  (TCM) \cite{Tavis1968} models.
Whereas the distribution of entangled states which is in the heart of quantum repeater is a main point in quantum information science and technologies, meanwhile
the influences of different types of dissipations including photon leakage from non-perfect cavity and spontaneous emission of atoms and \textit{etc} are unavoidable in these models \cite{Di2008}.
The existence of losses in the time evolution of entangled states generally may lead to the attenuation or even death of entanglement, \textit{i.e.}, the unwanted decoherence effects can be occurred.

 In \cite{Luong2016} the authors have proposed a scheme for overcoming losses by using quantum repeater. In \cite{Li2016} the quantum repeater with continuous variable encoding is discussed. Quantum repeater based on spatial entanglement of photons and quantum-dot spins in optical microcavities has been studied in \cite{Wang2012}. A model has been proposed  in  \cite{Van2009} for a long-line quantum repeater. Experimental demonstration of a quantum
 repeater node has been proposed in \cite{Yuan2008}, in which  two photons from two entangled atom-photon pairs overlap at BSM and the entanglement is generated between the two atomic ensembles. Recently, experimental schemes for quantum repeater have been reported \cite{Xu2017,Chen2017} wherein the authors have considered four entangled photon pairs $(1,2)$, $(3,4)$, $(5,6)$ and $(7,8)$, and studied the entanglement generation between photon pairs  $(1,4)$ and $(5,8)$  by performing BSM on photons $(2,3)$ and $(6,7)$. Finally, the entangled photons $(1,8)$ has been achieved by a further BSM on photons $(4,5)$. Now, in this paper a quantum repeater protocol is implemented via considering  a system including of eight aligned two-level atoms with ground state (excited state) $\ket{g}$ ($\ket{e}$) (see Fig. \ref{fig:Fig1}). Four pairs of  the qubits $(1,2), (3,4), (5,6), (7,8)$ are initially prepared in some atomic Bell states.  Our aim is to distribute entanglement between the two end atoms (1, 8) which are far from each other and possess a separable state. In particular, we want to consider the same system, however, to make the model much more close to experimental realization, we consider the dissipation sources namely atomic and field decay rates in the dynamical Hamiltonian and investigate their influences on the transferred entanglement to the atoms $(1,8)$.
 %
 %\hl{In experimental demonstration of quantum repeater} \cite{Xu2017,Chen2017} \hl{the fidelity is usually considered, but in this paper, we use the concurrence measure} \cite{Wootters1998} \hl{for calculating the entanglement, because in our discussion the final entangled states are converted to Bell or 	
 %specified entangled states.}
 Interestingly, as an important result of the present paper, we illustrate that in special conditions in which the two mentioned decay rates become equal, the dissipation effects automatically disappear from the entanglement dynamics of the system. Consequently, in this situation our considered dissipative quantum repeater is then free from the decoherence effects arisen from the two considered dissipation sources.\\
 This paper is organized as follows: We explain that how our quantum repeater protocol works in Sec. 2. The results of our investigations on the proposed protocol are analysed in Sec. 3. Finally,  the paper ends with summary and conclusions in Sec. 4.

%%%%%%%%%%%%%%%%%%%%%%%%%%%%%%%%%%%%%%%%%%%%%%%%%%%%
%===========================================================================
\section{Quantum repeater protocol}\label{model}
%===========================================================================
 \begin{figure}[H]
  \centering
     \subfigure[\label{fig.Fig1a} \ The BSM method]{\includegraphics[width=0.46\textwidth]{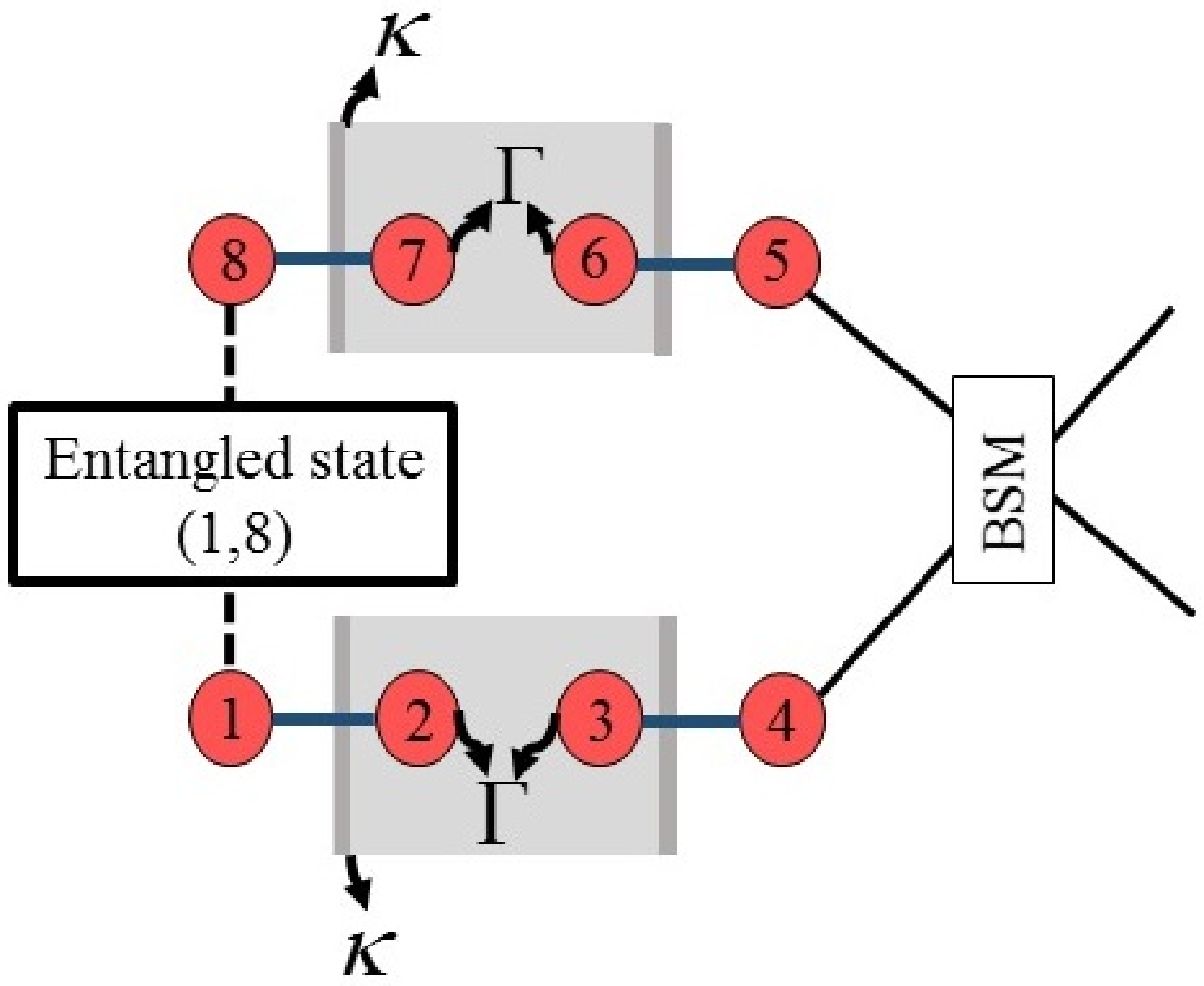}}
     \hspace{0.05\textwidth}
     \subfigure[\label{fig.Fig1b} \ The QED method]{\includegraphics[width=0.46\textwidth]{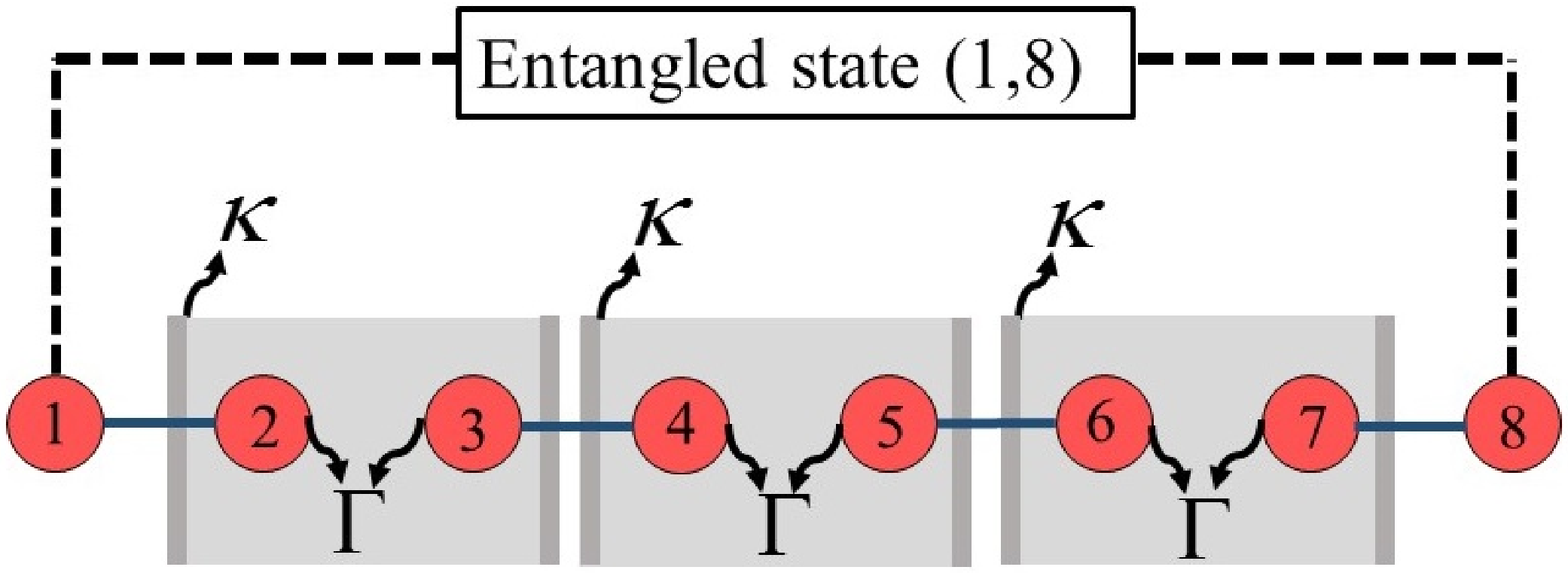}}

   \caption{\label{fig:Fig1} {The scheme of quantum repeater protocol in the presence of eight atoms $(1,2,...,8)$, where $\Gamma$ $(\kappa)$ is the spontaneous emission of atoms (photon leakage from the non-perfect cavities). The entangled atoms (1,8) are obtained by use of (a) BSM method (b) QED method.}}
  \end{figure}
 Our purpose in this paper is to distribute entanglement between two two-level atoms (1,8) among eight  aligned  atoms (1, 2, ..., 8) shown in Fig. \ref{fig:Fig1}. At first, our linearized system includes four entangled pairs of atoms ($i$,$i+1$) where $i=1,3,5,7$. Performing a dissipative  interaction  between the atoms (2, 3) and (6, 7) respectively results in the entanglement between atoms (1, 4) and (5, 8) \cite{Chun2006}. In the following, we demonstrate the mentioned  procedure for the produced entangled atomic states  (1, 4).
 The initial state of the atomic pairs (1, 2) and (3, 4) reads as $\ket{\Psi}_{1,2}\otimes\ket{\Psi}_{3,4}$, where
   \begin{eqnarray}\label{initialstate}
\ket{\Psi}_{i,i+1}&=&\frac{1}{\sqrt{2}}(\ket{e,g}-\ket{g,e})_{i,i+1}, \qquad   i=1,3.
  \end{eqnarray}
   Now, we perform the interaction between the two non-entangled qubits (2,3) possessing spontaneous emission rates ($\Gamma_2$, $\Gamma_3$) with a single-mode field in a non-perfect cavity with photon leakage rate ($\kappa$). The dynamics is governed by the following Hamiltonian\footnote[1]{\textcolor{blue}{The dissipative terms with $\kappa$ (rate of photon loss) and $\Gamma_i$ (rate of spontaneous emission) in Hamiltonian (\ref{hamiltonian}) have been considered in Refs. \cite{Ghasemi2017,Di2008,Fidio2009,Man2011,Grunwald2011} simultaneously using phenomenological method. Also, there are many references about this consideration such as Refs. \cite{Zhang2010,Baghshahi2016,Shore1993} for spontaneous emission and Ref. \cite{Barnett2007} for photon loss. Recall that, the rates of photon loss and spontaneous emission depend on the quality factor of cavity and the characteristics of atom, respectively.}} $(\hbar=1)$, $\hat{H}=\hat{H}_0+\hat{H}_1$,
  \begin{eqnarray}\label{hamiltonian}
  \hat{H}_0&=&\omega\hat{a}^{\dagger}\hat{a}+\textcolor{blue}{\sum_{i=2,3} \frac{\omega_i}{2}\hat{\sigma}_{iz} -\frac{ i}{2}\sum_{i=2,3} \Gamma_i\hat{\sigma}_i^+\hat{\sigma}_i^-} -i\frac{\kappa}{2}\hat{a}^{\dagger }\hat{a},  \\ \nonumber
  \hat{H}_1&=&g\sum_{i=2,3}\left( \hat{a}\hat{\sigma}_i^++\hat{a}^{\dagger}\hat{\sigma}_i^-\right),
  \end{eqnarray}
  where $\hat{H}_0$ ($\hat{H}_1$) is free (interacting) part of Hamiltonian in which $\hat{a}^{\dagger} (\hat{a})$ is creation (annihilation) operator, $\hat{\sigma}_i^+=\ket{e}_i\bra{g}(\hat{\sigma}_i^-=\hat{\sigma}_i^{+\dagger})$ and $\hat{\sigma}_{iz}$ are respectively the raising (lowering) and population inversion operators of the $i$th atom. Also, $\omega_i$ and $\omega$ are the $i$th atom transition and field frequencies, respectively.\\
  \textcolor{blue}{The interaction Hamiltonian in the interaction picture can be obtained via Baker-Hausdorff formula
   \begin{eqnarray}\label{hbl}
   \hat{H}_\mathrm{int}&=&e^{i \hat{H}_0 t} \hat{H}_1e^{-i \hat{H}_0 t}\\ \nonumber
   &=&\hat{H}_1+it\left[\hat{H}_0,\hat{H}_1 \right]+\frac{(it)^2}{2!} \left[\hat{H}_0,\left[\hat{H}_0,\hat{H}_1 \right]\right]+\cdots
    \end{eqnarray}
    where $\hat{H}_0$ and $\hat{H}_1$
  have been introduced in Eq. (\ref{hamiltonian}). Now, we have:
   \begin{eqnarray}\label{h}
   \left[\hat{H}_0,\hat{H}_1 \right]&=&g\left[ \omega_2-\omega+\frac{i}{2}(\kappa-\Gamma_2)\right]  \hat{a}\hat{\sigma}_2^+-g\left[ \omega_2-\omega+\frac{i}{2}(\kappa-\Gamma_2)\right]  \hat{a}^{\dagger}\hat{\sigma}_2^-\nonumber \\
   &+&g\left[ \omega_3-\omega+\frac{i}{2}(\kappa-\Gamma_3)\right]  \hat{a}\hat{\sigma}_3^+-g\left[ \omega_3-\omega+\frac{i}{2}(\kappa-\Gamma_3)\right]  \hat{a}^{\dagger}\hat{\sigma}_3^-,
   \end{eqnarray}
   and
   \begin{eqnarray}\label{h2}
   \left[\hat{H}_0,[\hat{H}_0,\hat{H}_1 ] \right]&=&g\left[ \omega_2-\omega+\frac{i}{2}(\kappa-\Gamma_2)\right]^2  \hat{a}\hat{\sigma}_2^++g\left[ \omega_2-\omega+\frac{i}{2}(\kappa-\Gamma_2)\right]^2  \hat{a}^{\dagger}\hat{\sigma}_2^-\nonumber \\
    &+&g\left[ \omega_3-\omega+\frac{i}{2}(\kappa-\Gamma_3)\right]^2  \hat{a}\hat{\sigma}_3^++g\left[ \omega_3-\omega+\frac{i}{2}(\kappa-\Gamma_3)\right]^2  \hat{a}^{\dagger}\hat{\sigma}_3^-.
   \end{eqnarray}
   By substituting Eqs. (\ref{h}), (\ref{h2}) into the formula (\ref{hbl}), one can obtain the interaction Hamiltonian in the interaction picture as (for $\Delta_2=\Delta_3=\Delta$ where $\Delta_i=\omega_i-\omega$ and $\Gamma_2=\Gamma_3=\Gamma$):
  \begin{eqnarray}\label{iphamiltonian}
  \hat{H}_{int}&=&g\sum_{i=2,3}\left( \hat{a}\hat{\sigma}_i^+ e^{i\delta t}+\hat{a}^{\dagger}\hat{\sigma}_i^- e^{-i\delta t}\right) ,\\
  \delta&=&\Delta+\frac{i}{2}(\kappa-\Gamma).\nonumber
  \end{eqnarray}}
   \textcolor{blue}{Following the path of Ref. \cite{Zheng2000}, we assume that the cavity is in vacuum state, so the effective Hamiltonian under the limit of large detuning, \textit{i.e.}, $\lvert \delta \lvert \gg \lvert g \lvert$, is achieved as below (in the case $\lvert \delta \lvert \gg \lvert g \lvert$, there is no energy exchange between
   the atomic system and the cavity \cite{Zheng2000}):
\begin{eqnarray}\label{effectivehamiltonian}
\hat{H}_{eff}&=&\lambda\sum_{i=2,3}\hat{\sigma}_i^+\hat{\sigma}_i^-+ \lambda\left(\hat{\sigma}_2^+\hat{\sigma}_3^-+H.C.\right) ,
\end{eqnarray}}
  \textcolor{blue}{with $\lambda=\frac{g^2}{\delta}$. It is readily found that, in the above Hamiltonians (\ref{iphamiltonian}), (\ref{effectivehamiltonian})} if the cavity and atomic decay rates are chosen such that $\kappa=\Gamma$, the influence of dissipation is then removed.
 Finally, by using the effective Hamiltonian (\ref{effectivehamiltonian}) and initial state $\ket{\Psi}_{1,2}\otimes\ket{\Psi}_{3,4}$ (see Eq. (\ref{initialstate})) the entangled state for atoms $(1-4)$ is achieved as follows:
\begin{eqnarray}\label{state1-4}
 \ket{\Psi(t)}_{1,2,3,4}&=&\frac{1}{N(t)}(L_1(t)\ket{eegg}+L_2(t)\ket{egeg}+L_3(t)\ket{egge}\\ \nonumber
 &+&L_4(t)\ket{geeg} + L_5(t)\ket{gege}+L_6(t)\ket{ggee})_{1,2,3,4},
\end{eqnarray}
 where
\begin{eqnarray}\label{coefficient}
  L_1(t)&=&L_6(t)=-i\frac{e^{-i\lambda t}}{2}\sin{\lambda t},\\ \nonumber
  L_2(t)&=&L_5(t)=\frac{e^{-i\lambda t}}{2} \cos{\lambda t},\\ \nonumber
  L_3(t)&=&-\frac{1}{2},\qquad   L_4(t)=-\frac{e^{-i2\lambda t}}{2},\\ \nonumber
  N(t)&=&\sqrt{\sum_{i=1}^{6} \left| L_i(t)\right|^2}.
 \end{eqnarray}
   Now, if one applies a measurement on the atomic state $(2, 3)$ and obtains  $\ket{eg}_{2,3}$ or $\ket{ge}_{2,3}$, the outcome results for atoms $(1, 4)$ are  respectively as follow:
 \begin{eqnarray}\label{state14}
    \ket{\Psi(t)}_{1,4}&=&\frac{1}{\sqrt{\left| L_1(t)\right|^2+\left| L_5(t)\right|^2} }(L_1(t)\ket{eg}+L_5(t)\ket{ge})_{1,4},\nonumber \\
    \ket{\Psi'(t)}_{1,4}&=&\frac{1}{\sqrt{\left| L_2(t)\right|^2+\left| L_6(t)\right|^2} }(L_2(t)\ket{eg}+L_6(t)\ket{ge})_{1,4}.
 \end{eqnarray}
 It is clearly observed that the qubit pair $(1,4)$ is now entangled.\\
 In a similar manner, the above procedure can be  repeated for atoms $(5-8)$, so that the following states can readily be achieved for the previously non-entangled pair (5,8) as below:
\begin{eqnarray}\label{state58}
   \ket{\Psi(t)}_{5,8}&=&\frac{1}{\sqrt{\left| L_1(t)\right|^2+\left| L_5(t)\right|^2} }(L_1(t)\ket{eg}+L_5(t)\ket{ge})_{5,8},\nonumber \\
   \ket{\Psi'(t)}_{5,8}&=&\frac{1}{\sqrt{\left| L_2(t)\right|^2+\left| L_6(t)\right|^2} }(L_2(t)\ket{eg}+L_6(t)\ket{ge})_{5,8}.
\end{eqnarray}
Now, there exist four possible states in our hand, which constitute from atoms $(1,4,5,8)$, \textit{i.e.}, $ \ket{\Psi(t)}_{1,4}\otimes \ket{\Psi(t)}_{5,8}$, $ \ket{\Psi(t)}_{1,4}\otimes \ket{\Psi'(t)}_{5,8}$, $ \ket{\Psi'(t)}_{1,4}\otimes \ket{\Psi(t)}_{5,8}$ and $ \ket{\Psi'(t)}_{1,4}\otimes \ket{\Psi'(t)}_{5,8}$.  The entanglement is now can be swapped  from atoms (1,4) and (5,8) into the atoms (1,8) by the BSM method (Fig. \ref{fig.Fig1a}) or again performing dissipative interaction between atoms (4,5) \textcolor{blue}{with the interaction time from $t$ to $\tau$} (QED method) (Fig. \ref{fig.Fig1b}). In the following subsections we consider all possible cases in detail.

\subsection[The first state]{The state $ \ket{\Psi(t)}_{1,4}\otimes \ket{\Psi(t)}_{5,8}$}
The operation of a BSM  using the  Bell state
$\ket{B}_{4,5}=\frac{1}{\sqrt{2}}(\ket{ee}+\ket{gg})_{4,5}$ \cite{Ai2007} on the state $ \ket{\Psi(t)}_{1,4}\otimes \ket{\Psi(t)}_{5,8}$,
the state of atoms $(1,8)$ is converted to the Bell state
\begin{equation}\label{bellstate}
\ket{B}_{1,8}=\frac{1}{\sqrt{2}}(\ket{ee}+\ket{gg})_{1,8},
\end{equation}
with the success probability
\begin{equation}\label{sucb}
S_{\ket{B}_{1,8}}(t)=\frac{\left| L_1(t) L_5(t) \right|^2 }{(\left|L_1(t) \right|^2+ \left|L_5(t) \right|^2)^2}.
\end{equation}
We continue our process via performing a BSM with the Bell state $\ket{B'}_{4,5}=\frac{1}{\sqrt{2}}(\ket{eg}+\ket{ge})_{4,5}$ \cite{Ai2007}
 on $ \ket{\Psi(t)}_{1,4}\otimes \ket{\Psi(t)}_{5,8}$. Consequently the state of atoms (1,8) is transferred to the following state
\begin{equation}
 \ket{\gamma}^1_{1,8}=\frac{1}{\sqrt{ \left| L_1(t) \right|^4+ \left| L_5(t) \right|^4}}(L^2_1(t)\ket{eg}+L^2_5(t)\ket{ge})_{1,8},
\end{equation}
whose concurrence \cite{Wootters1998} may be straightforwardly obtained as
\begin{equation}
C_1(t)=\frac{2\left| L^{*2}_1(t) L^2_5(t) \right| }{\left|L_1(t) \right|^4+ \left|L_5(t) \right|^4}.
\end{equation}
In this status the entangled atoms (1,8) are produced by performing the dissipative interaction under the introduced dynamics in (\ref{effectivehamiltonian}) between atoms (4,5).
The unnormalized entangled state for atoms $(1,4,5,8)$ after interaction with initial state $ \ket{\Psi(t)}_{1,4}\otimes \ket{\Psi(t)}_{5,8}$ is achieved as

\begin{equation}\label{intstate1458}
\begin{aligned}
\small
   \ket{\phi}_{1,4,5,8} & = \frac{1}{\left| L_1(t) \right|^2+\left| L_5(t) \right|^2}\left\lbrace L^2_1(t)\left[\frac{1}{2}(e^{i2\lambda t} e^{-i2\lambda \tau}-1) \ket{eegg}\right.\right.  \\
        &+ \left. \left. \frac{1}{2}(e^{i2\lambda t} e^{-i2\lambda \tau}+1)\ket{egeg}\right]\right.
          +\left.  L^2_5(t) \left[\frac{1}{2}(e^{i2\lambda t} e^{-i2\lambda \tau}+1) \ket{gege}\right. \right. \\
          & +\left.  \frac{1}{2}(e^{i2\lambda t} e^{-i2\lambda \tau}-1)\ket{ggee} \right] +  L_1(t)  L_5(t) \left(\ket{egge}\right. \\
             & +\left. \left.   e^{-i2\lambda (\tau-t)} \ket{geeg} \right) \right\rbrace _{1,4,5,8},
\end{aligned}
\end{equation}
\textcolor{blue}{where $\tau$ is the time of interaction between atoms (4,5) from $t$ to $\tau$.} By measuring the state $\ket{eg}_{4, 5}$ on state (\ref{intstate1458}), the entangled state of atoms (1,8) collapses to
 \begin{eqnarray}\label{intstate18}
    \ket{\phi}^1_{1,8}&= &\frac{1}{\sqrt{\left| a_1(t,\tau)\right|^2+\left| b_1(t,\tau)\right|^2 }}\left(a_1(t,\tau)\ket{eg}+b_1(t,\tau)\ket{ge} \right) _{1,8},\\ \nonumber
 a_1(t,\tau)&=&\frac{L^2_1(t)}{2}(e^{i2\lambda t} e^{-i2\lambda \tau}-1),\\ \nonumber
 b_1(t,\tau)&=&\frac{L^2_5(t)}{2}(e^{i2\lambda t} e^{-i2\lambda \tau}+1),
  \end{eqnarray}
which possesses the concurrence as
\begin{equation}
C'_1(t,\tau)= \frac{2\left|a^*_1(t,\tau) b_1(t,\tau) \right| }{\left| a_1(t,\tau)\right|^2+\left| b_1(t,\tau)\right|^2 }.
\end{equation}
   Moreover, by applying a measurement on the state (\ref{intstate1458}) to obtain the atomic state $(4, 5)$ as $\ket{ge}_{4, 5}$,  the following entangled state for atoms $(1,8)$ is obtained
     \begin{eqnarray}\label{intstatee18}
        \ket{\phi'}^1_{1,8}&= &\frac{1}{\sqrt{\left| a'_1(t,\tau)\right|^2+\left| b'_1(t,\tau)\right|^2 }}\left(a'_1(t,\tau)\ket{eg}+b'_1(t,\tau)\ket{ge} \right) _{1,8},\\ \nonumber
     a'_1(t,\tau)&=&\frac{L^2_1(t)}{2}(e^{i2\lambda t} e^{-i2\lambda \tau}+1),\\ \nonumber
     b'_1(t,\tau)&=&\frac{L^2_5(t)}{2}(e^{i2\lambda t} e^{-i2\lambda \tau}-1),
      \end{eqnarray}
   with the corresponding concurrence as
   \begin{equation}
C^{\prime\prime}_1(t,\tau)= \frac{2\left|a'^*_1(t,\tau) b'_1(t,\tau) \right| }{\left| a'_1(t,\tau)\right|^2+\left| b'_1(t,\tau)\right|^2 }.
   \end{equation}

 \subsection[The second state]{The state $ \ket{\Psi(t)}_{1,4}\otimes \ket{\Psi'(t)}_{5,8}$}

By the operation of BSM with the state  $\ket{B'}_{4,5}=\frac{1}{\sqrt{2}}(\ket{eg}+\ket{ge})_{4,5}$ on the state $ \ket{\Psi(t)}_{1,4}\otimes \ket{\Psi'(t)}_{5,8}$, the following Bell state
 \begin{equation}\label{belll}
\ket{B'}_{1,8}=\frac{1}{\sqrt{2}}(\ket{eg}+\ket{ge})_{1,8},
 \end{equation}
   is achieved for atoms (1, 8) with the success probability
   \begin{equation}\label{sucbp}
   S_{\ket{B'}_{1,8}}(t)=\frac{\left| L_1(t) L_2(t) \right|^2 +\left| L_5(t) L_6(t) \right|^2}{2 (\left|L_1(t) \right|^2+ \left|L_5(t) \right|^2)^2}.
   \end{equation}
 Notice that $S_{\ket{B}_{1,8}}(t)=S_{\ket{B'}_{1,8}}(t)$. Also, by the operation of a BSM using the Bell state $\ket{B}_{4,5}=\frac{1}{\sqrt{2}}(\ket{ee}+\ket{gg})_{4,5}$ on the state $\ket{\Psi(t)}_{1,4}\otimes \ket{\Psi'(t)}_{5,8}$, the state of atoms (1, 8) is transferred to the entangled state as below:
   \begin{equation}
\ket{\gamma}^2_{1,8}=\frac{1}{\sqrt{ \left| L_1(t) L_6(t) \right|^2+ \left| L_5(t) L_2(t) \right|^2}}(L_1(t) L_6(t)\ket{ee}+ L_2(t) L_5(t)\ket{gg})_{1,8},
   \end{equation}
and concurrence is calculated as follows:
\begin{equation}
C_2(t)=\frac{2\left| L^*_1(t) L^*_6(t) L_5(t) L_2(t) \right| }{ \left| L_1(t) L_6(t) \right|^2+ \left| L_5(t) L_2(t) \right|^2}.
\end{equation}
 Notice that $C_2(t)=C_1(t)$. But, by performing the dissipative interaction using the Hamiltonian (\ref{effectivehamiltonian}) between atoms $(4,5)$ with initial state $ \ket{\Psi(t)}_{1,4}\otimes \ket{\Psi'(t)}_{5,8}$ the unnormalized entangled state for atoms $(1,4,5,8)$ is achieved as
 \begin{equation}\label{intstate1458e}
 \begin{aligned}
    \ket{\phi'}_{1,4,5,8} & = \frac{1}{\left| L_1(t) \right|^2+\left| L_5(t) \right|^2}\left\lbrace L_1(t) L_2(t)\left[\frac{1}{2}(e^{i2\lambda t} e^{-i2\lambda \tau}-1) \ket{eegg}\right. \right. \\
    &+\left.  \frac{1}{2}(e^{i2\lambda t} e^{-i2\lambda \tau}+1)\ket{egeg}\right] + L_5(t)  L_6(t) \left[\frac{1}{2} (e^{i2\lambda t} e^{-i2\lambda \tau}+1)\ket{gege}\right. \\
          &+\left. \left. \frac{1}{2}(e^{i2\lambda t} e^{-i2\lambda \tau}-1) \ket{ggee} \right]
        + L_1(t)L_6(t)  \ket{egge}\right. \\
        &+\left. e^{-2i\lambda (\tau-t)} L_2(t)L_5(t)\ket{geeg}  \right\rbrace _{1,4,5,8}.
 \end{aligned}
 \end{equation}
 Now, by measuring the states $\ket{eg}_{4,5}$ and $\ket{ge}_{4,5}$ on state (\ref{intstate1458e}), after normalization we arrive respectively at the states of atoms (1,8) as,
  \begin{eqnarray}\label{intstate182}
     \ket{\phi}^2_{1,8}&= &\frac{1}{\sqrt{\left| a_2(t,\tau)\right|^2+\left| b_2(t,\tau)\right|^2 }}\left(a_2(t,\tau)\ket{eg}+b_2(t,\tau)\ket{ge} \right) _{1,8},\\ \nonumber
  a_2(t,\tau)&=&\frac{L_1(t)L_2(t)}{2}(e^{i2\lambda t} e^{-i2\lambda \tau}-1),\\ \nonumber
  b_2(t,\tau)&=&\frac{L_5(t)L_6(t)}{2}(e^{i2\lambda t} e^{-i2\lambda \tau}+1),
   \end{eqnarray}
   with concurrence
   \begin{equation}\label{c2}
 C'_2(t,\tau)=\frac{2\left|a^*_2(t,\tau) b_2(t,\tau) \right| }{\left| a_2(t,\tau)\right|^2+\left| b_2(t,\tau)\right|^2 },
   \end{equation}
   and
      \begin{eqnarray}\label{intstatee182}
          \ket{\phi'}^2_{1,8}&= &\frac{1}{\sqrt{\left| a'_2(t,\tau)\right|^2+\left| b'_2(t,\tau)\right|^2 }}\left(a'_2(t,\tau)\ket{eg}+b'_2(t,\tau)\ket{ge} \right) _{1,8},\\ \nonumber
       a'_2(t,\tau)&=&\frac{L_1(t)L_2(t)}{2}(e^{i2\lambda t} e^{-i2\lambda \tau}+1),\\ \nonumber
       b'_2(t,\tau)&=&\frac{L_5(t)L_6(t)}{2}(e^{i2\lambda t} e^{-i2\lambda \tau}-1),
        \end{eqnarray}
 with concurrence
         \begin{eqnarray}\label{concuu2}
           C^{\prime\prime}_2(t,\tau)&= &\frac{2\left|a'^*_2(t,\tau) b'_2(t,\tau) \right| }{\left| a'_2(t,\tau)\right|^2+\left| b'_2(t,\tau)\right|^2 }.
          \end{eqnarray}
 It is easy to check that $ C^{\prime\prime}_2(t,\tau)=C'_2(t,\tau)$.
    \subsection[The third state]{The state $ \ket{\Psi'(t)}_{1,4}\otimes \ket{\Psi(t)}_{5,8}$}
 By performing BSM with $\ket{B'}_{4,5}=\frac{1}{\sqrt{2}}(\ket{eg}+\ket{ge})_{4,5}$ on $ \ket{\Psi'(t)}_{1,4}\otimes \ket{\Psi(t)}_{5,8}$, the entangled state of atoms $(1,8)$ is transferred to Bell state $\ket{B'}_{1,8}$ in Eq. (\ref{belll}) with success probability (\ref{sucbp}).
  But, by performing the BSM with the help of Bell state $\ket{B}_{4,5}=\frac{1}{\sqrt{2}}(\ket{ee}+\ket{gg})_{4,5}$ on the state $ \ket{\Psi'(t)}_{1,4}\otimes \ket{\Psi(t)}_{5,8}$, the state of atoms (1,8) is converted to
  \begin{equation}
\ket{\gamma}^3_{1,8}=\frac{1}{\sqrt{ \left| L_2(t)  L_5(t) \right|^2+\left| L_1(t)  L_6(t) \right|^2}}(L_2(t) L_5(t)\ket{ee}+L_1(t) L_6(t)\ket{gg})_{1,8},
  \end{equation}
with the concurrence as
       \begin{equation}
     C_3(t)=\frac{2\left| L^*_2(t) L^*_5(t) L_1(t) L_6(t) \right| }{\left|L_2(t)L_5(t) \right|^2+ \left|L_1(t)L_6(t) \right|^2}.
       \end{equation}
       Notice that $C_3(t)=C_2(t)=C_1(t)$. The entangled state of atoms $(1,4,5,8)$ is achieved by performing the interaction (\ref{effectivehamiltonian}) between atoms $(4,5)$ with initial state $ \ket{\Psi'(t)}_{1,4}\otimes \ket{\Psi(t)}_{5,8}$ as below,
   \begin{equation}\label{intstate1458ee}
   \begin{aligned}
      \ket{\phi^{\prime\prime}}_{1,4,5,8} & = \frac{1}{\left| L_1(t) \right|^2+\left| L_5(t) \right|^2}\left\lbrace L_1(t) L_2(t)\left[\frac{1}{2}(e^{i2\lambda t} e^{-i2\lambda \tau}-1) \ket{eegg}\right. \right. \\
      &+ \left.  \frac{1}{2}(e^{i2\lambda t} e^{-i2\lambda \tau}+1)\ket{egeg}\right]+  L_5(t)  L_6(t) \left[ \frac{1}{2}(e^{i2\lambda t} e^{-i2\lambda \tau}+1)\ket{gege}\right. \\
      &+\left.  \frac{1}{2}(e^{i2\lambda t} e^{-i2\lambda \tau}-1) \ket{ggee} \right]
            + L_2(t)L_5(t)  \ket{egge}\\
            &+\left. e^{-2i\lambda (\tau-t)} L_1(t)L_6(t)\ket{geeg}  \right\rbrace _{1,4,5,8}.
   \end{aligned}
   \end{equation}
       Now, by measuring the states $\ket{eg}_{4,5}$ and $\ket{ge}_{4,5}$ on state (\ref{intstate1458ee}), the entangled states of atoms $(1,8)$ are respectively converted to $\ket{\phi}^3_{1,8}$ and $\ket{\phi'}^3_{1,8}$, where $\ket{\phi}^3_{1,8}= \ket{\phi}^2_{1,8}$ and $\ket{\phi'}^3_{1,8}= \ket{\phi'}^2_{1,8}$ in Eqs. (\ref{intstate182}) and (\ref{intstatee182}), with calculated concurrences $C'_3(t,\tau)=C'_2(t,\tau)$ and $C^{\prime\prime}_3(t,\tau)=C^{\prime\prime}_2(t,\tau)$ as in Eqs. (\ref{c2}) and (\ref{concuu2}), respectively. It is easy to see that $C'_3(t,\tau)=C'_2(t,\tau)=C^{\prime\prime}_2(t,\tau)=C^{\prime\prime}_3(t,\tau)$.

   \subsection[The forth state]{The state $ \ket{\Psi'(t)}_{1,4}\otimes \ket{\Psi'(t)}_{5,8}$}
   We apply the BSM using $\ket{B}_{4,5}=\frac{1}{\sqrt{2}}(\ket{ee}+\ket{gg})_{4,5}$ on state $ \ket{\Psi'(t)}_{1,4}\otimes \ket{\Psi'(t)}_{5,8}$, so the state of atoms (1,8) is converted to Bell state $\ket{B}_{1,8}$ in Eq. (\ref{bellstate}) with the success probability
   \begin{equation}\label{sucbpp}
   S'_{\ket{B}_{1,8}}(t)=\frac{\left| L_2(t) L_6(t) \right|^2 }{(\left|L_1(t) \right|^2+ \left|L_5(t) \right|^2)^2}.
   \end{equation}
   Notice that $S'_{\ket{B}_{1,8}}(t)=S_{\ket{B'}_{1,8}}(t)=S_{\ket{B}_{1,8}}(t)$. Also, by the operation of BSM method with the Bell state $\ket{B'}_{4,5}=\frac{1}{\sqrt{2}}(\ket{eg}+\ket{ge})_{4,5}$ on state $ \ket{\Psi'(t)}_{1,4}\otimes \ket{\Psi'(t)}_{5,8}$, the state of atoms (1,8) reads as
   \begin{equation}
  \ket{\gamma}^4_{1,8}=\frac{1}{\sqrt{ \left| L_2(t) \right|^4+ \left| L_6(t) \right|^4}}(L^2_2(t)\ket{eg}+L^2_6(t)\ket{ge})_{1,8},
   \end{equation}
    with the concurrence
    \begin{equation}
C_4(t)=\frac{2\left| L^{*2}_2(t) L^2_6(t) \right| }{\left|L_2(t) \right|^4+ \left|L_6(t) \right|^4}.
    \end{equation}
    It can straightforwardly be checked that, $C_4(t)=C_3(t)=C_2(t)=C_1(t)$.
    The unnormalized entangled state of atoms $(1,4,5,8)$ is obtained by performing dissipative interaction (Eq. (\ref{effectivehamiltonian})) with initial state $ \ket{\Psi'(t)}_{1,4}\otimes \ket{\Psi'(t)}_{5,8}$ as,
  \begin{equation}\label{intstate14584}
  \begin{aligned}
  \small
     \ket{\phi^{\prime\prime\prime}}_{1,4,5,8} & =\frac{1}{\left| L_2(t) \right|^2+\left| L_6(t) \right|^2} \left\lbrace L^2_2(t)\left[\frac{1}{2}(e^{i2\lambda t} e^{-i2\lambda \tau}-1) \ket{eegg}\right. \right. \\
     &+\left.  \frac{1}{2}(e^{i2\lambda t} e^{-i2\lambda \tau}+1)\ket{egeg}\right]+  L^2_6(t) \left[\frac{1}{2}(e^{i2\lambda t} e^{-i2\lambda \tau}+1) \ket{gege}\right. \\
     &+\left. \frac{1}{2}(e^{i2\lambda t} e^{-i2\lambda \tau}-1)\ket{ggee} \right] +  L_2(t)  L_6(t) \left(\ket{egge}\right. \\
           &+\left. \left. e^{-i2\lambda (\tau-t)} \ket{geeg} \right) \right\rbrace _{1,4,5,8}.
  \end{aligned}
 \end{equation}
   By measuring the state $\ket{eg}_{4,5}$ on state (\ref{intstate14584}), the state of atoms (1,8) collapses to,
   \begin{eqnarray}\label{intstate184}
      \ket{\phi}^4_{1,8}&= &\frac{1}{\sqrt{\left| a_4(t,\tau)\right|^2+\left| b_4(t,\tau)\right|^2 }}\left(a_4(t,\tau)\ket{eg}+b_4(t,\tau)\ket{ge} \right) _{1,8},\\ \nonumber
   a_4(t,\tau)&=&\frac{L^2_2(t)}{2}(e^{i2\lambda t} e^{-i2\lambda \tau}-1),\\ \nonumber
   b_4(t,\tau)&=&\frac{L^2_6(t)}{2}(e^{i2\lambda t} e^{-i2\lambda \tau}+1),
    \end{eqnarray}
  with the concurrence,
  \begin{equation}
 C'_4(t,\tau)=\frac{2\left|a^*_4(t,\tau) b_4(t,\tau) \right| }{\left| a_4(t,\tau)\right|^2+\left| b_4(t,\tau)\right|^2 }.
  \end{equation}
Notice that, $ C'_4(t,\tau)=C^{\prime\prime}_1(t,\tau)$. Also, by measuring the state $\ket{ge}_{4,5}$ on state (\ref{intstate14584}), the state of atoms (1,8) is converted to the entangled state
   \begin{eqnarray}\label{intstatee184}
      \ket{\phi'}^4_{1,8}&= &\frac{1}{\sqrt{\left| a'_4(t,\tau)\right|^2+\left| b'_4(t,\tau)\right|^2 }}\left(a'_4(t,\tau)\ket{eg}+b'_4(t,\tau)\ket{ge} \right) _{1,8},\\ \nonumber
   a'_4(t,\tau)&=&\frac{L^2_2(t)}{2}(e^{i2\lambda t} e^{-i2\lambda \tau}+1),\\ \nonumber
   b'_4(t,\tau)&=&\frac{L^2_6(t)}{2}(e^{i2\lambda t} e^{-i2\lambda \tau}-1),
    \end{eqnarray}
  whose concurrence reads as
  \begin{equation}
C^{\prime\prime}_4(t,\tau)= \frac{2\left|a'^*_4(t,\tau) b'_4(t,\tau) \right| }{\left| a'_4(t,\tau)\right|^2+\left| b'_4(t,\tau)\right|^2 }.
  \end{equation}
  It is clearly seen that $C^{\prime\prime}_4(t,\tau)=C'_1(t,\tau)$.
 \section{Results and discussion} \label{sec.results}
 \begin{figure}[H]
 \centering
 \subfigure[\label{fig.Fig2a} \ $C_1(t)=C_2(t)=C_3(t)=C_4(t)$]{\includegraphics[width=0.7\textwidth]{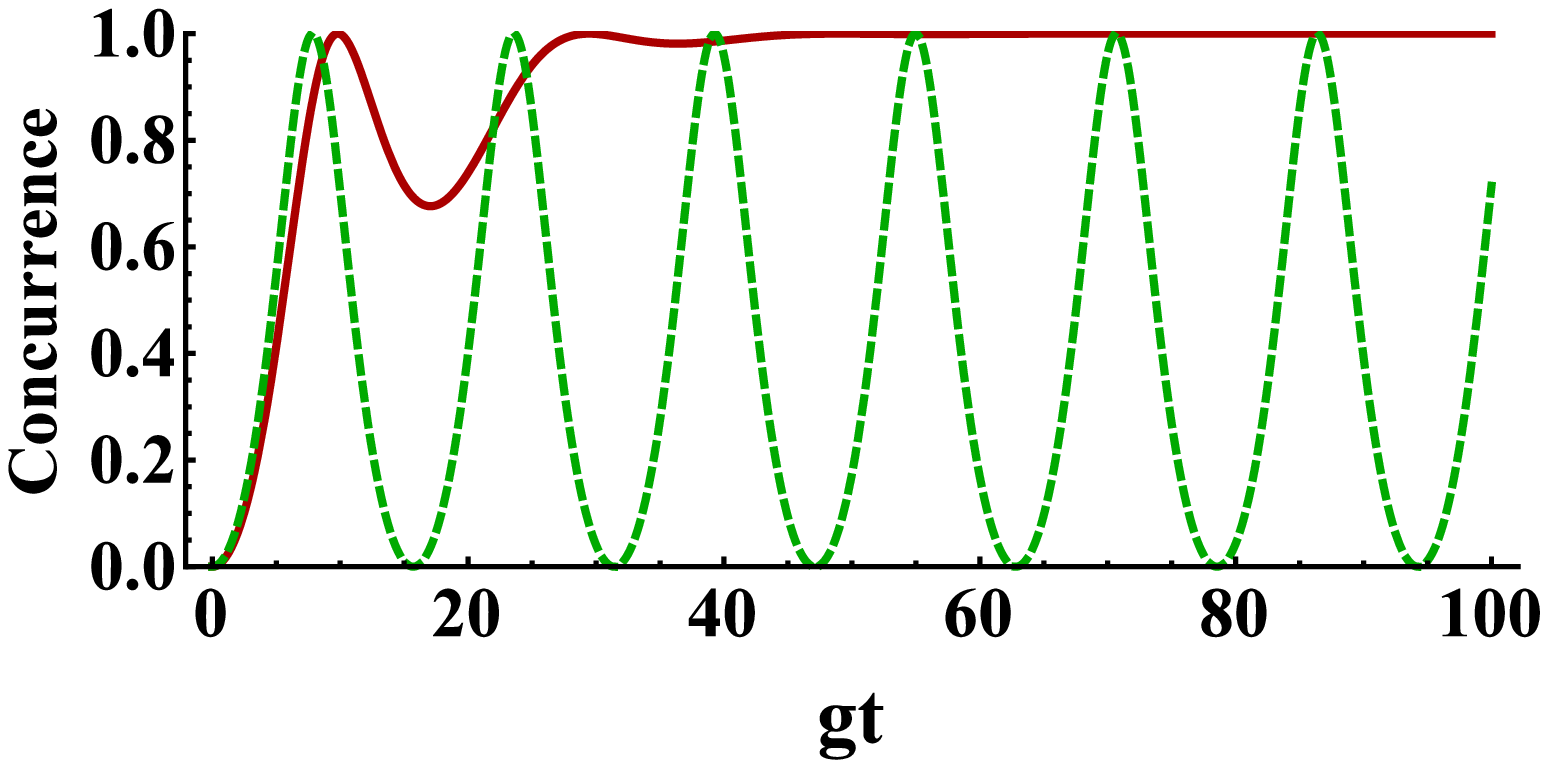}}
 \hspace{0.05\textwidth}
 \subfigure[\label{fig.Fig2b} \ $C_1(t)=C_2(t)=C_3(t)=C_4(t)$]{\includegraphics[width=0.7\textwidth]{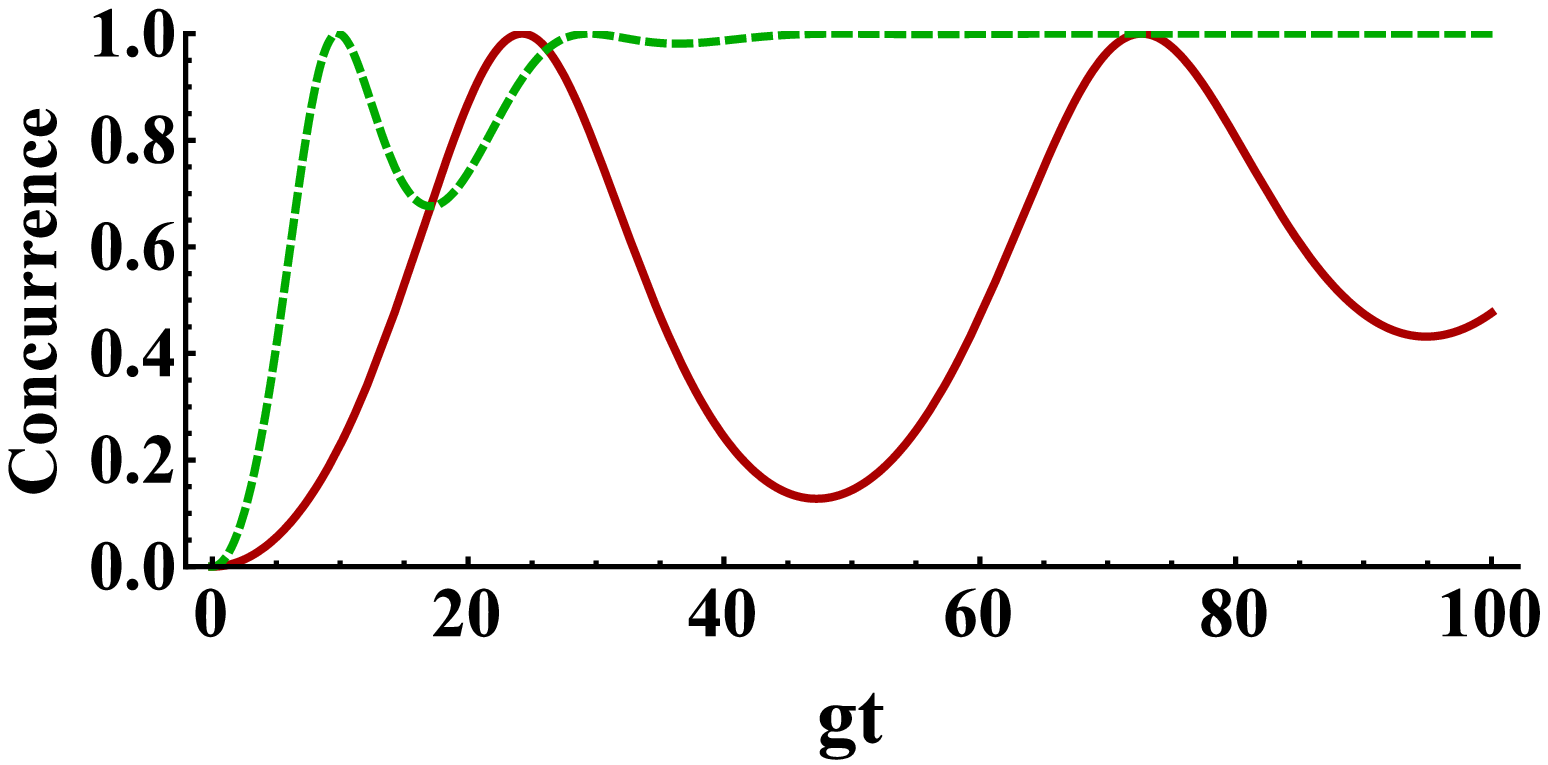}}
 \caption{\label{fig.cdissdet} {\it The effect of dissipation and detuning on the evolution of concurrence}: (a) $C_1(t)=C_2(t)=C_3(t)=C_4(t)$ for $\kappa=\Gamma$ (dashed green line) and \textcolor{blue}{$\kappa=20g, \Gamma=10g$} (solid red line) with $\Delta=10g$ and (b) $C_1(t)=C_2(t)=C_3(t)=C_4(t)$  for $\Delta=10g$ (dashed green line) and $\Delta=30g$ (solid red line) with $\kappa=20g$ and $\Gamma=10g$.}
 \end{figure}
 \begin{figure}[H]
  \centering
  \subfigure[\label{fig.Fig3a} \ $\small{S'_{\ket{B}_{1,8}}(t)=S_{\ket{B'}_{1,8}}(t)=S_{\ket{B}_{1,8}}(t)}$]{\includegraphics[width=0.7\textwidth]{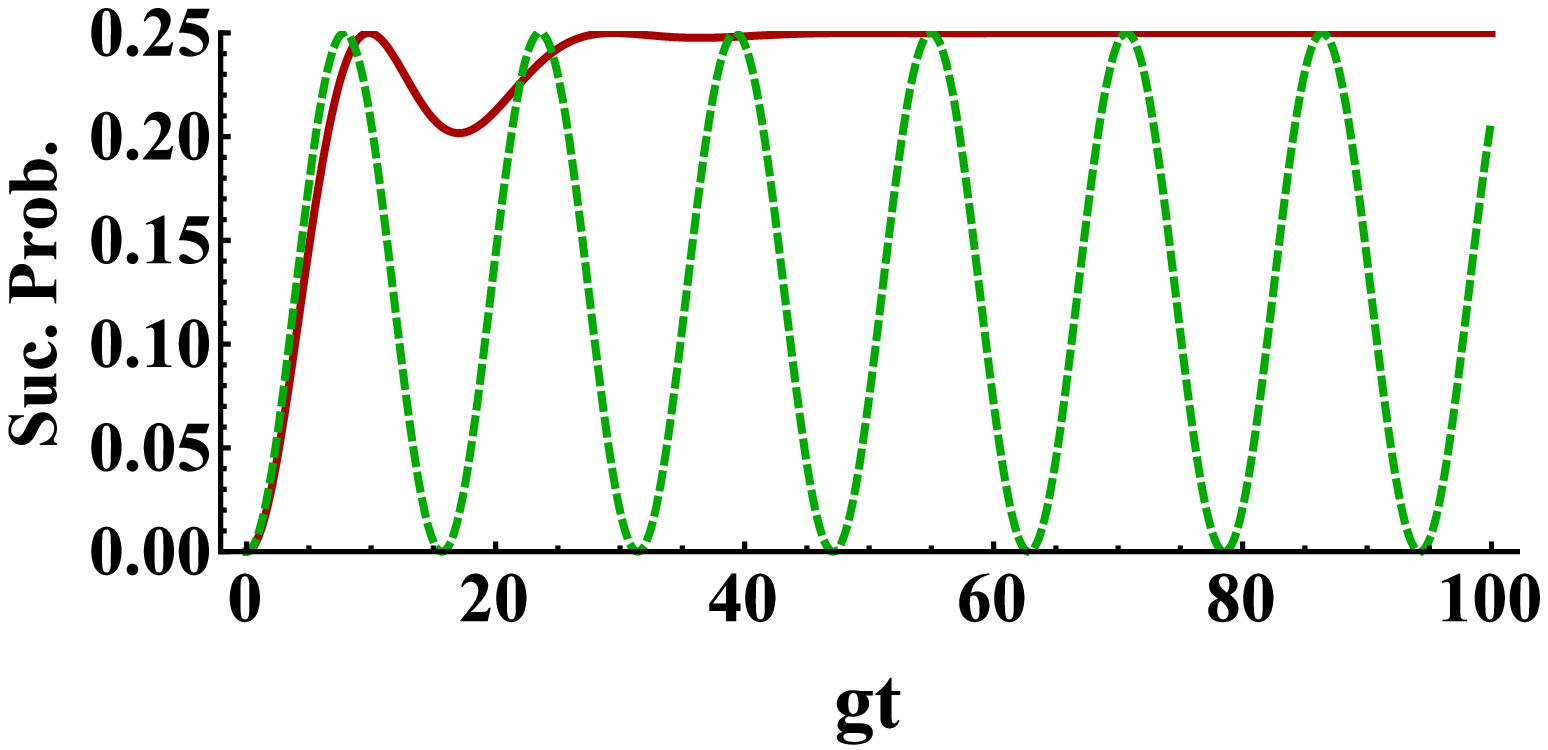}}
  \hspace{0.05\textwidth}
  \subfigure[\label{fig.Fig3b} \ $\small{S'_{\ket{B}_{1,8}}(t)=S_{\ket{B'}_{1,8}}(t)=S_{\ket{B}_{1,8}}(t)}$]{\includegraphics[width=0.7\textwidth]{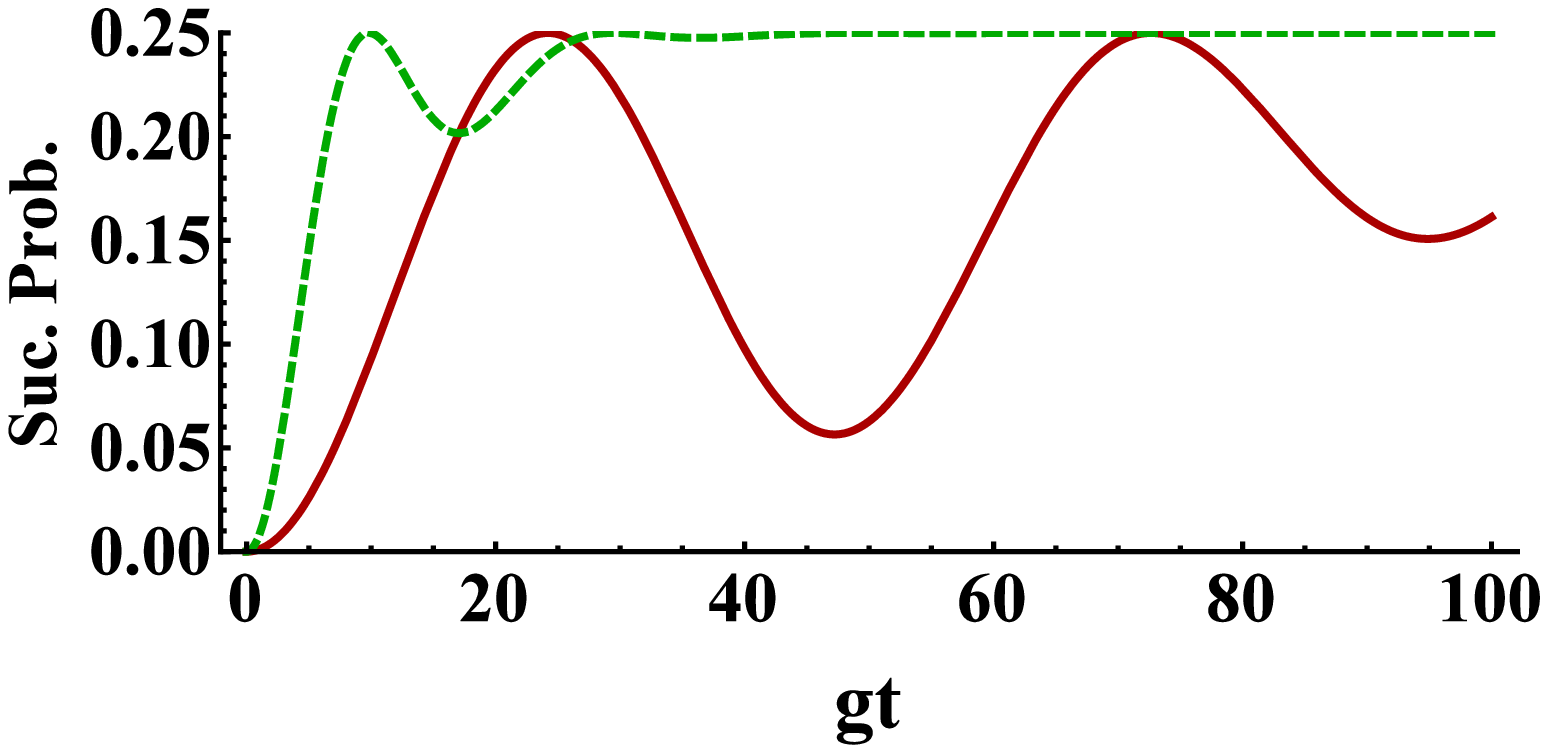}}
  \caption{\label{fig.sdissdet} {\it The effect of dissipation and detuning on the evolution of success probability}: (a) $S'_{\ket{B}_{1,8}}(t)=S_{\ket{B'}_{1,8}}(t)=S_{\ket{B}_{1,8}}(t)$, for $\kappa=\Gamma$ (dashed green line) and \textcolor{blue}{$\kappa=20g, \Gamma=10g$} (solid red line) with $\Delta=10g$ and (b) $\small{S'_{\ket{B}_{1,8}}(t)=S_{\ket{B'}_{1,8}}(t)=S_{\ket{B}_{1,8}}(t)}$, for $\Delta=10g$ (dashed green line) and $\Delta=30g$ (solid red line) with $\kappa=20g$ and $\Gamma=10g$.}
  \end{figure}
\begin{figure}[H]
 \centering
 \subfigure[\label{fig.Fig4a} \ $C'_1(t,\tau)=C^{\prime\prime}_4(t,\tau)$]{\includegraphics[width=0.7\textwidth]{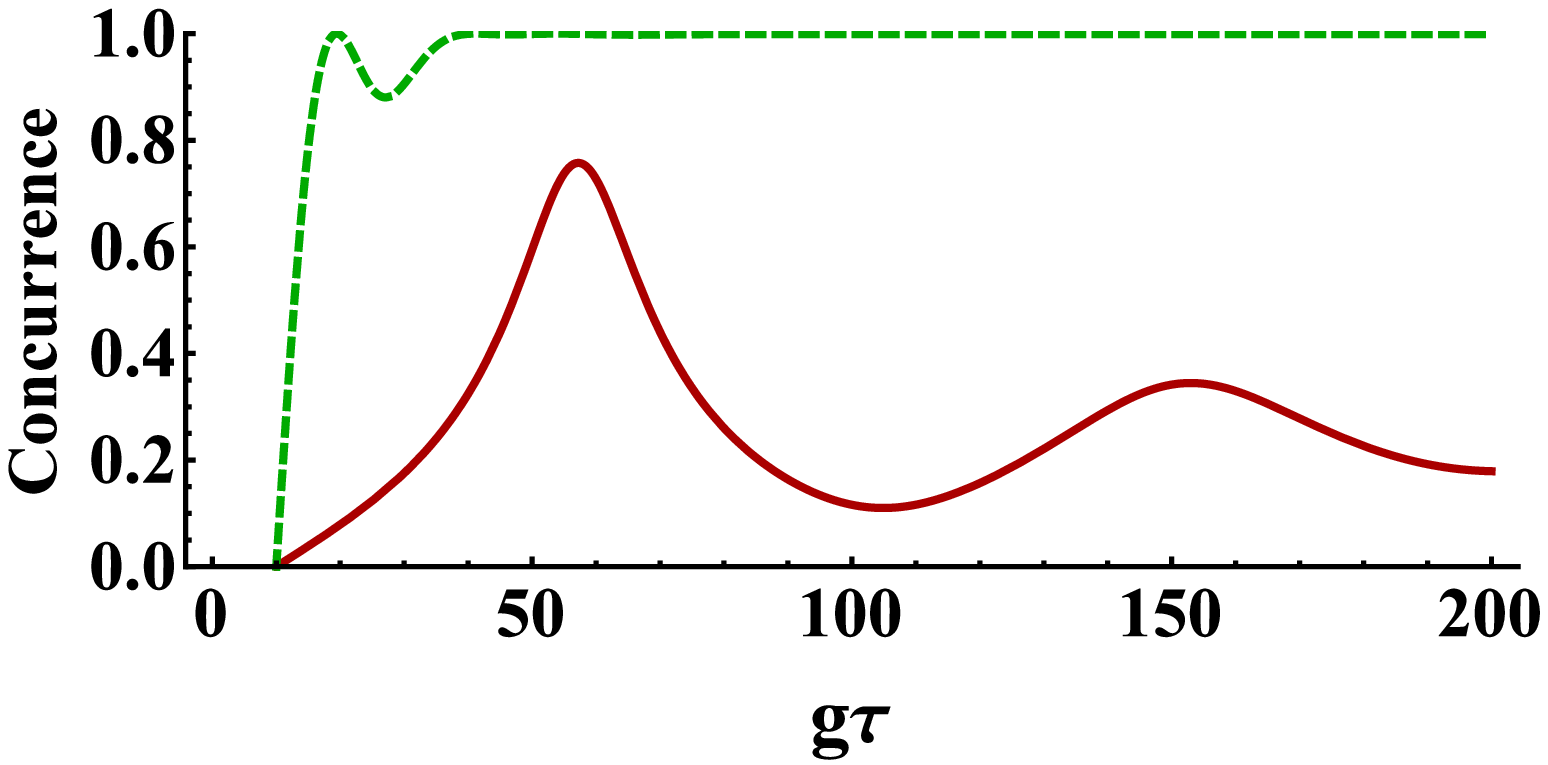}}
 \hspace{0.05\textwidth}
 \subfigure[\label{fig.Fig4b} \ $C^{\prime\prime}_1(t,\tau)=C'_4(t,\tau)$]{\includegraphics[width=0.7\textwidth]{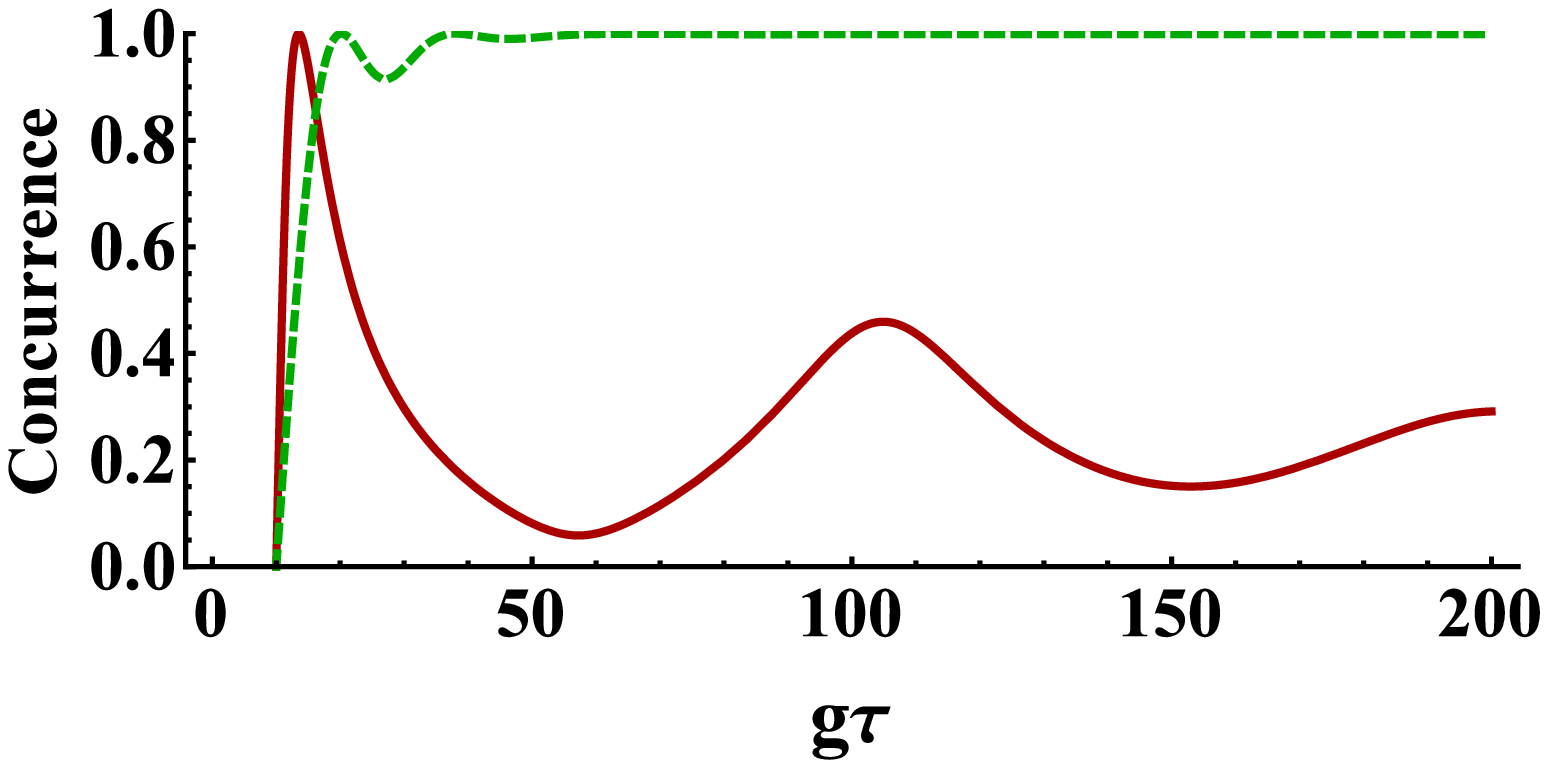}}
  \hspace{0.05\textwidth}
  \subfigure[\label{fig.Fig4c} \ $\small{C'_2(t,\tau)=C^{\prime\prime}_2(t,\tau)=C'_3(t,\tau)=C^{\prime\prime}_3(t,\tau)}$]{\includegraphics[width=0.7\textwidth]{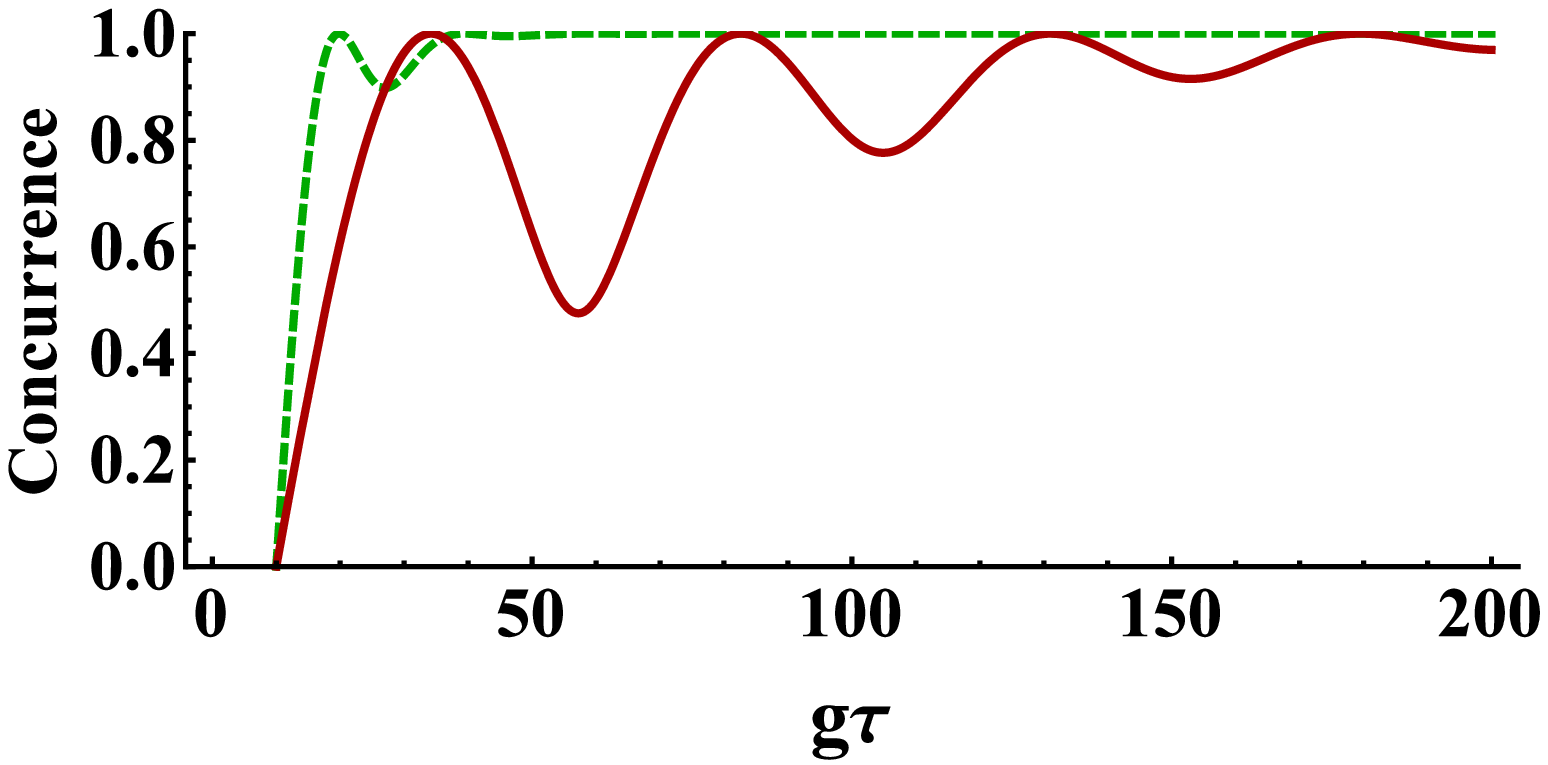}}

 \caption{\label{fig.cdet} {\it The effect of detuning on the evolution of concurrence}: (a) $C'_1(t,\tau)=C^{\prime\prime}_4(t,\tau)$ (b) $C^{\prime\prime}_1(t,\tau)=C'_4(t,\tau)$ (c)  $C'_2(t,\tau)=C^{\prime\prime}_2(t,\tau)=C'_3(t,\tau)=C^{\prime\prime}_3(t,\tau)$, for $\Delta=10g$ (dashed green line) and $\Delta=30g$ (solid red line) with \textcolor{blue}{$\kappa=20g, \Gamma=10g$} and $gt=10$.}
  \end{figure}
   \begin{figure}[H]
   \centering
   \subfigure[\label{fig.Fig5a} \ $C'_1(t,\tau)=C^{\prime\prime}_4(t,\tau)$]{\includegraphics[width=0.7\textwidth]{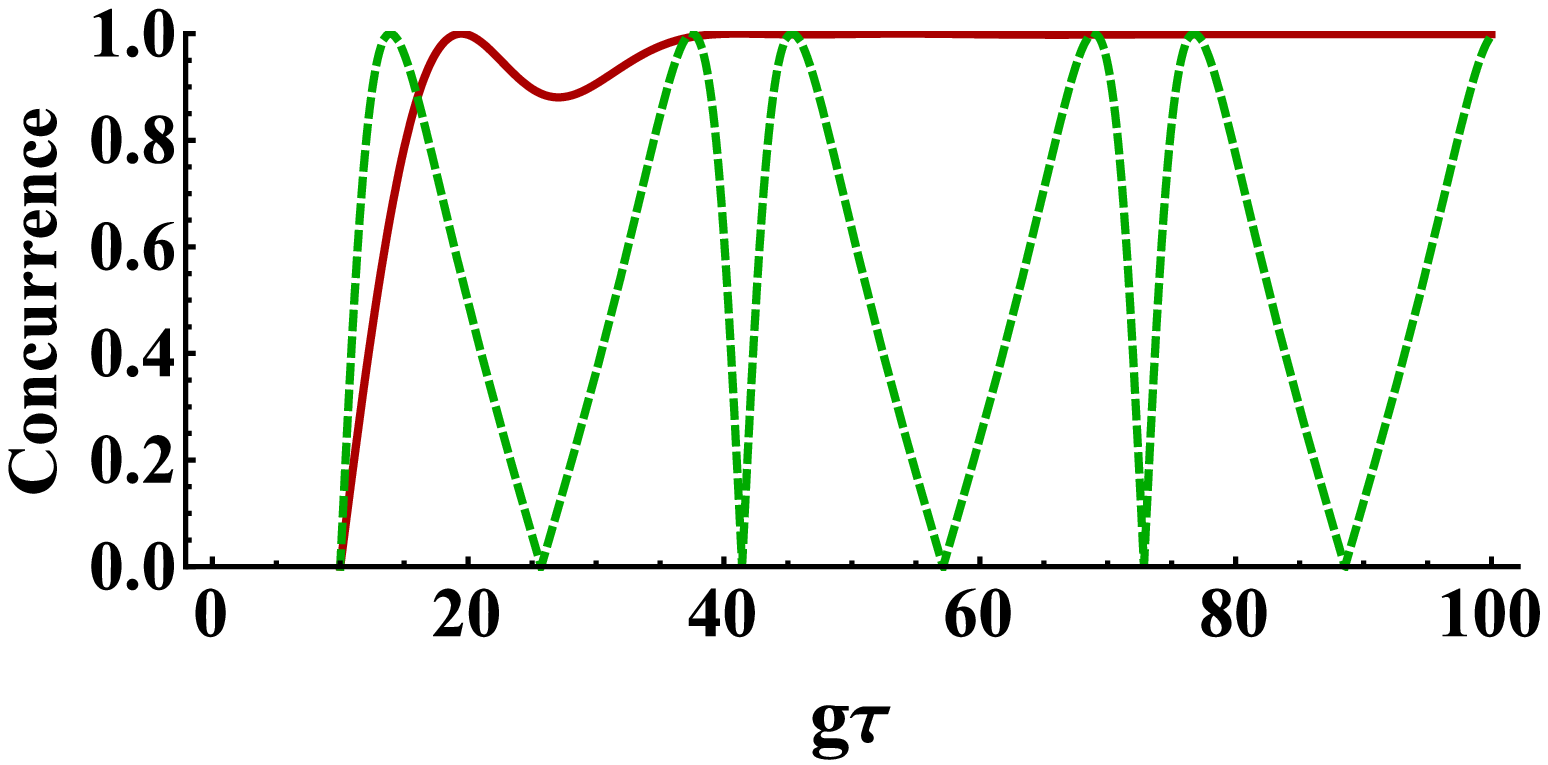}}
   \hspace{0.05\textwidth}
   \subfigure[\label{fig.Fig5b} \ $C^{\prime\prime}_1(t,\tau)=C'_4(t,\tau)$]{\includegraphics[width=0.7\textwidth]{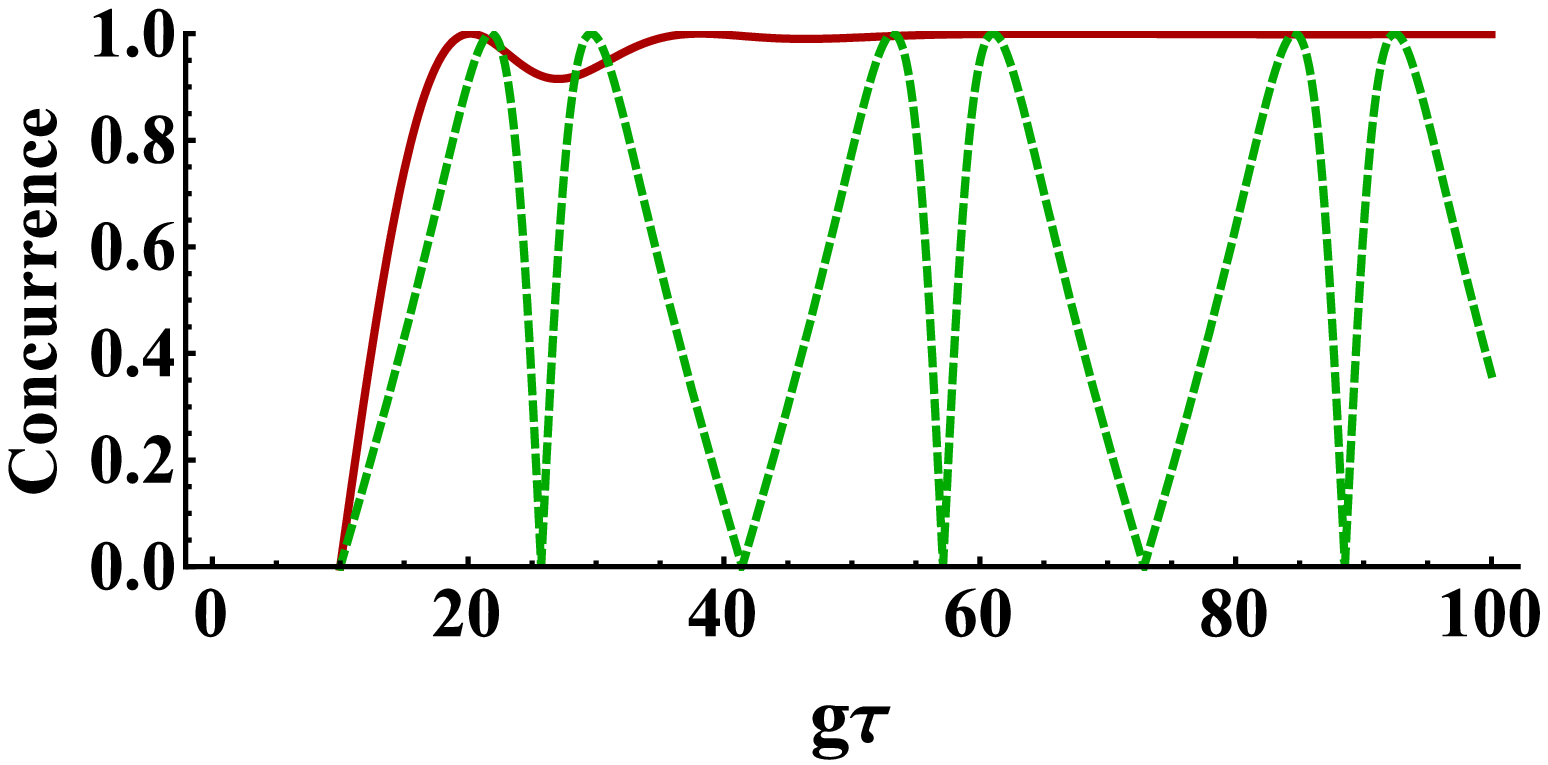}}
      \hspace{0.05\textwidth}
   \subfigure[\label{fig.Fig5c} \ $\small{C'_2(t,\tau)=C^{\prime\prime}_2(t,\tau)=C'_3(t,\tau)=C^{\prime\prime}_3(t,\tau)}$]{\includegraphics[width=0.7\textwidth]{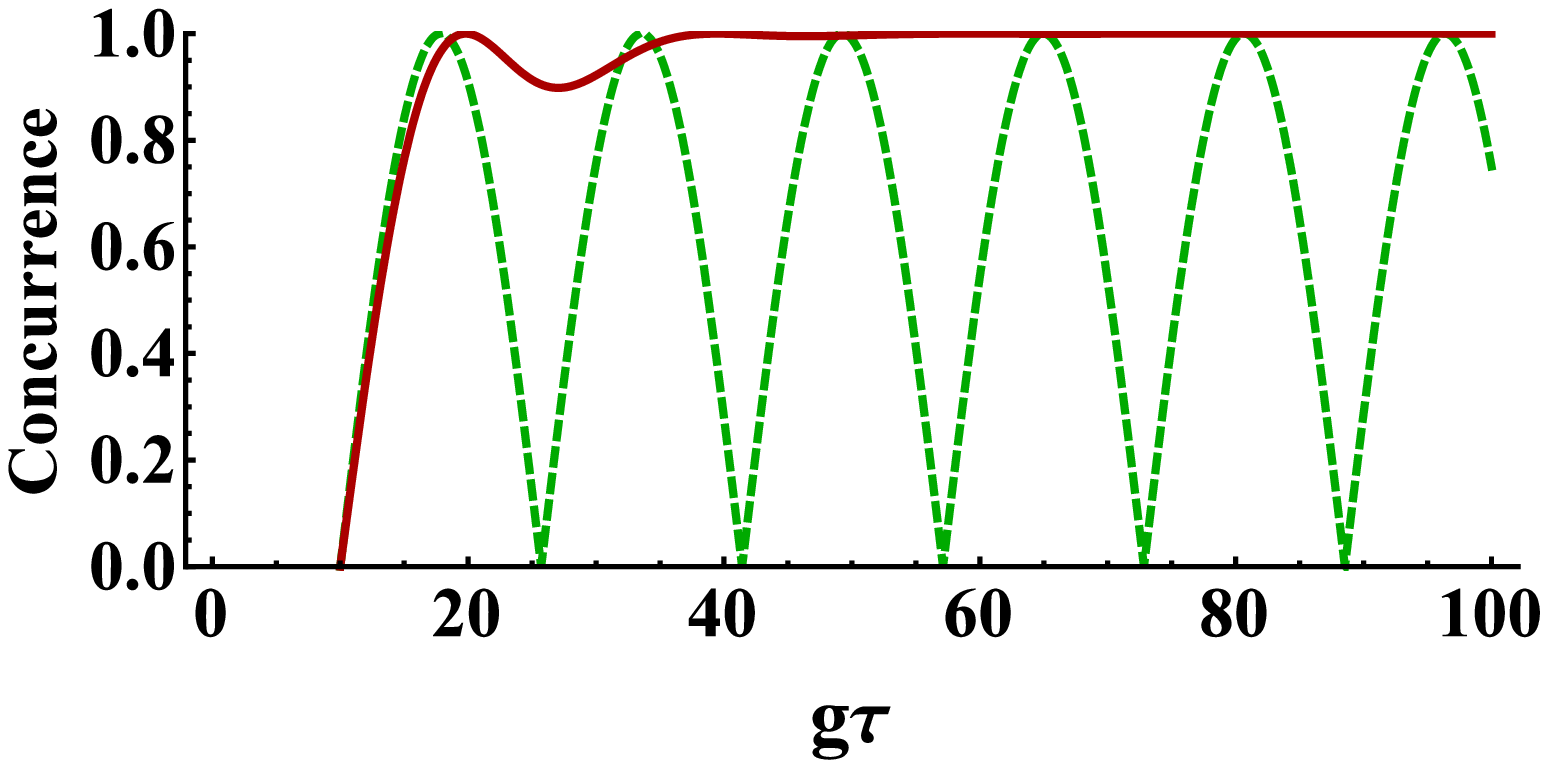}}

   \caption{\label{fig.cdiss} {\it The effect of dissipation on the evolution of concurrence}: (a) $C'_1(t,\tau)=C^{\prime\prime}_4(t,\tau)$ (b) $C^{\prime\prime}_1(t,\tau)=C'_4(t,\tau)$ (c)  $C'_2(t,\tau)=C^{\prime\prime}_2(t,\tau)=C'_3(t,\tau)=C^{\prime\prime}_3(t,\tau)$, for $\kappa=\Gamma$ (dashed green line) and \textcolor{blue}{$\kappa=20g, \Gamma=10g$} (solid red line) with $\Delta=10g$ and $gt=10$.}
   \end{figure}
      \begin{figure}[H]
      \centering
      \subfigure[\label{fig.Fig6a} \ $C'_1(t,\tau)=C^{\prime\prime}_4(t,\tau)$]{\includegraphics[width=0.7\textwidth]{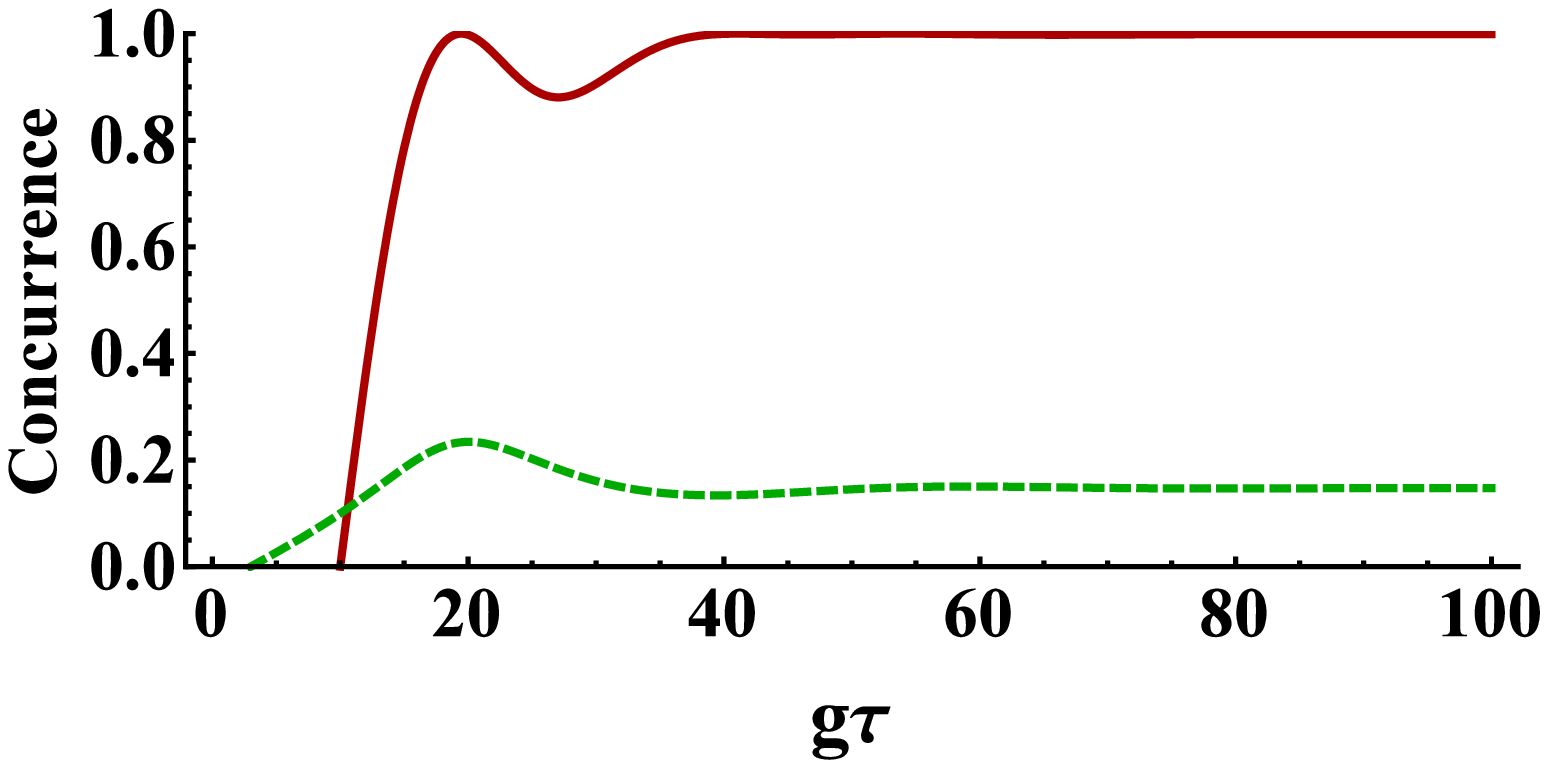}}
      \hspace{0.05\textwidth}
      \subfigure[\label{fig.Fig6b} \ $C^{\prime\prime}_1(t,\tau)=C'_4(t,\tau)$]{\includegraphics[width=0.7\textwidth]{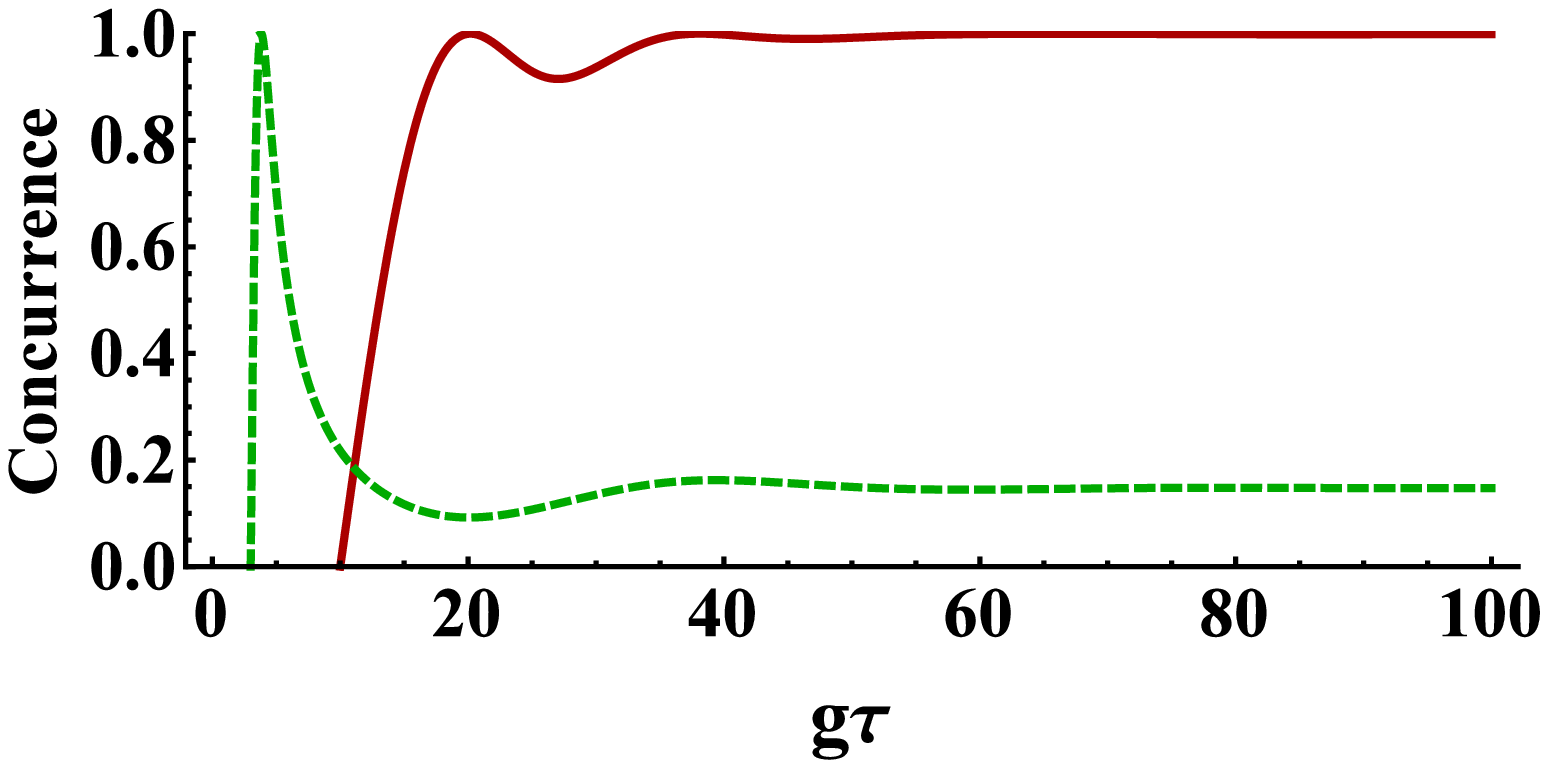}}
         \hspace{0.05\textwidth}
      \subfigure[\label{fig.Fig6c} \ $\small{C'_2(t,\tau)=C^{\prime\prime}_2(t,\tau)=C'_3(t,\tau)=C^{\prime\prime}_3(t,\tau)}$]{\includegraphics[width=0.7\textwidth]{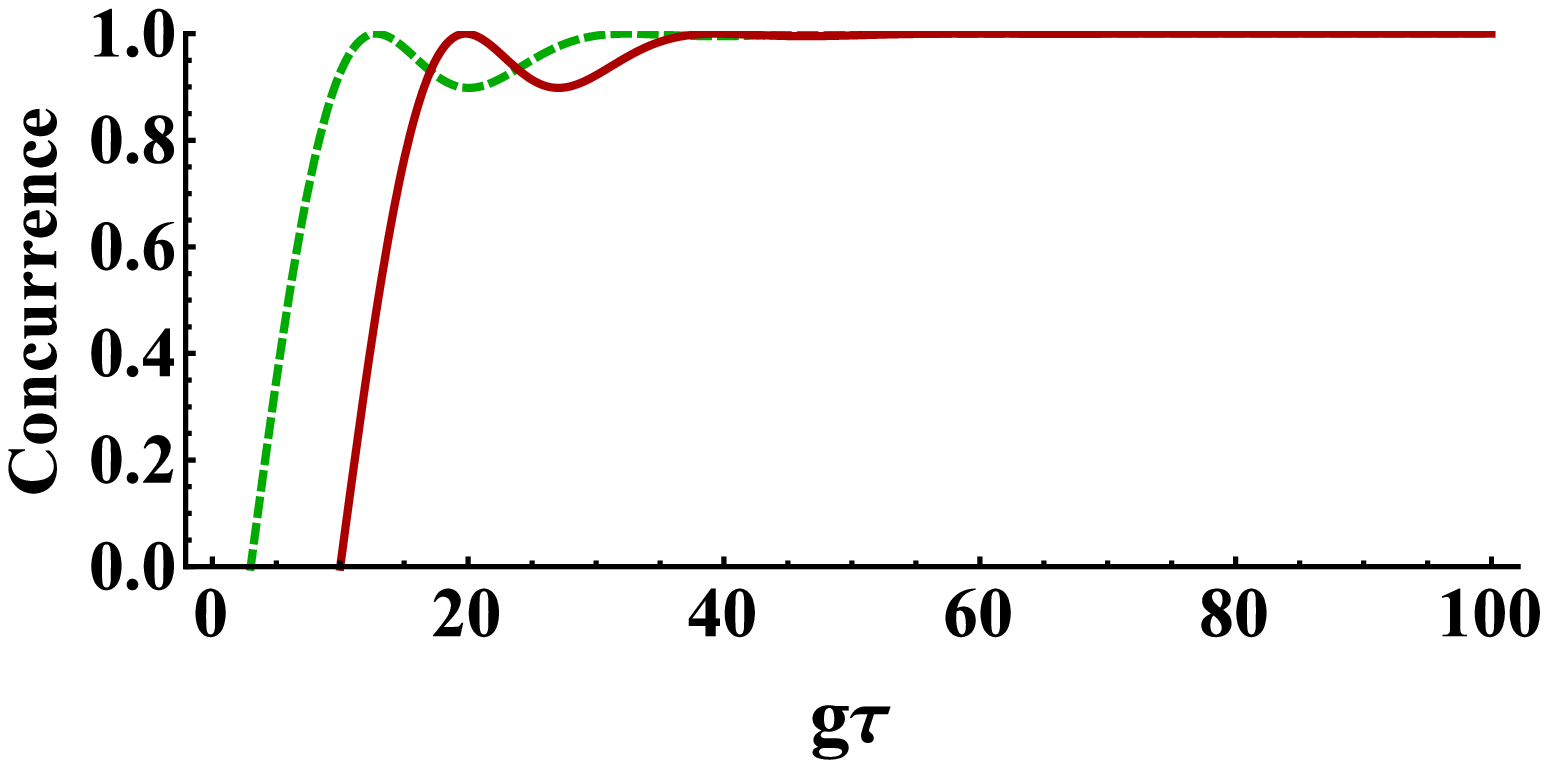}}

      \caption{\label{fig.ct} {\it The effect of initial state (which depends on choosing $gt$) on the evolution of concurrence}: (a) $C'_1(t,\tau)=C^{\prime\prime}_4(t,\tau)$ (b) $C^{\prime\prime}_1(t,\tau)=C'_4(t,\tau)$ (c) $C'_2(t,\tau)=C^{\prime\prime}_2(t,\tau)=C'_3(t,\tau)=C^{\prime\prime}_3(t,\tau)$, for $gt=3$ (dashed green line) and $gt=10$ (solid red line) with $\Delta=10g$, \textcolor{blue}{$\kappa=20g$ and $\Gamma=10g$}.}
      \end{figure}
 In this section we have analysed the effects of dissipation and detuning on the time evolution of concurrence or entanglement swapped between two atoms $(1,8)$ and also success probability of maximally entangled states produced by BSM method. Also, we have considered the effects of detuning, dissipation and initial interaction time which is entered by using different values of $gt$ on the concurrence obtained in QED method. The effects of dissipation and detuning on the degree of entanglement (success probability) calculated in BSM method have been shown in Fig. \ref{fig.cdissdet} (\ref{fig.sdissdet}), \textit{i.e.}, the regular evolution of concurrence (success probability) in the absence of dissipation  \textcolor{blue}{(the effect of dissipation in Eq. (\ref{iphamiltonian}) is removed by choosing $\kappa=\Gamma$)} in Fig.  \ref{fig.Fig2a} (\ref{fig.Fig3a}) is observed. But, evolution of concurrence (success probability) in the presence of dissipation after one oscillation reaches to $1$ ($0.25$) and remains at this maximum so that sudden death of entanglement has been removed in Fig. {\ref{fig.Fig2a}}. In Fig. \ref{fig.Fig2b} (\ref{fig.Fig3b}) the effect of detuning on the concurrence (success probability) has been considered in the presence of dissipation  \textcolor{blue}{$(\kappa\neq\Gamma)$}. We can see that the concurrence (success probability) has been reached to its maximum and remains at this value after an oscillation by decreasing the detuning. \textcolor{blue}{Generally, in Fig. \ref{fig.cdissdet} the state of target atoms (1,8) is converted to a maximally entangled state in some times with acceptable success probability shown in Fig. \ref{fig.sdissdet}}.\\ In Fig. \ref{fig.cdet} the effect of detuning on concurrence evaluated in QED method has been considered in the presence of dissipation \textcolor{blue}{$(\kappa\neq\Gamma)$}. \textcolor{blue}{In Fig. \ref{fig.cdet} solid red lines have been plotted for increased detuning and these lines are compared with the dashed green lines.} The concurrence in Fig. \ref{fig.Fig4a} has not been reached to its maximum even though detuning is increased. In Fig. \ref{fig.Fig4b} the maximum of concurrence has been achieved once for increased detuning. Also, in Fig. \ref{fig.Fig4c} the amplitude of oscillation has been decreased as time goes on by increasing the detuning. From Fig. \ref{fig.cdet} one can see that, decreasing the detuning causes the stability of entanglement and the maximum of concurrences is obtained in a large time interval.\\ In Fig. \ref{fig.cdiss} we have considered the effect of dissipation on the concurrence obtained by QED method. \textcolor{blue}{In this figure the dashed green lines (solid red lines) have been plotted in the absence (presence) of dissipation.} We can see that the periodic evolution of concurrence in the absence of dissipation $(\kappa=\Gamma)$ and this regular behaviour is destroyed in the presence of dissipation $(\kappa\neq\Gamma)$. Also, in the presence of dissipation the atoms (1,8) have been converted to atomic Bell states and remain at these states after an oscillation of concurrence in Figs. \ref{fig.Fig5a}, \ref{fig.Fig5b} and \ref{fig.Fig5c}.\\ The interaction time $gt$ \textcolor{blue}{between atoms (2,3) and (6,7)} and its importance have been considered in Fig. \ref{fig.ct}. \textcolor{blue}{In Fig. \ref{fig.ct} solid red lines have been plotted for increased interaction time and these lines are compared with the dashed green lines. In Figs. \ref{fig.Fig6a}, \ref{fig.Fig6b} and \ref{fig.Fig6c} by increasing the initial time $gt$, the states of atoms (1,8) have been converted to Bell states after one oscillation. So, the interaction time between atoms (2,3) and (6,7) is very effective in order to achieve the stable maximally entangled states for atoms (1,8)}
\section{Summary and conclusions} \label{sec.Conclusion}
Our purpose in this paper was the distribution of entanglement between two distant two-level atoms that have no interaction between each other. By using the quantum repeater protocol we contrasted with entanglement attenuation in our considered dissipative protocol. We supposed that eight atoms located in four maximally entangled atomic states (1,2), (3,4), (5,6) and (7,8). At first, by performing interaction between atoms (2,3) and (6,7), separately, in the presence of spontaneous emission and photon leakage, the entanglement was produced between atoms (1,4) and (5,8). In this level, we swapped entanglement from entangled atoms (1,4) and (5,8) to atoms (1,8) by BSM and QED methods. The atoms (1,8) converted to maximally entangled states by choosing suitable Bell states with the help of BSM method. We calculated the entanglement and success probability for atoms (1,8) and we found that the time evolutions of these parameters are periodic in the absence of dissipation. But, in the presence of dissipation this regular behaviour is destroyed. Interestingly for particular case $\kappa=\Gamma$, our system converted to ideal system with no dissipation. This situation may be achieved by choosing a suitable cavity and a particular two-level atom. We considered the effects of detuning, dissipation and interaction time $gt$ (\textit{i.e.}, the initial state) on entanglement dynamics. We saw the destructive effects of increasing (decreasing) the detuning (interaction time $gt$) on the concurrence. The concurrence was reached to its maximum after an oscillation in a large time interval by decreasing (increasing) the detuning (interaction time) in the presence of dissipation. Also, sudden death of entanglement was removed in the presence of dissipation and the stability of entanglement was achieved for concurrences.
%\newpage
\bibliographystyle{spmpsci}%
\bibliography{ref}

\begin{thebibliography}{10}
\providecommand{\url}[1]{{#1}}
\providecommand{\urlprefix}{URL }
\expandafter\ifx\csname urlstyle\endcsname\relax
  \providecommand{\doi}[1]{DOI~\discretionary{}{}{}#1}\else
  \providecommand{\doi}{DOI~\discretionary{}{}{}\begingroup
  \urlstyle{rm}\Url}\fi

\bibitem{Briegel1998}
Briegel, H.J., D{\"u}r, W., Cirac, J.I., Zoller, P.: Quantum repeaters: the
  role of imperfect local operations in quantum communication.
\newblock Phys. Rev. Lett. \textbf{81}(26), 5932 (1998)

\bibitem{Dur1999}
D{\"u}r, W., Briegel, H.J., Cirac, J., Zoller, P.: Erratum: Quantum repeaters
  based on entanglement purification.
\newblock Phys. Rev. A \textbf{60}(1), 725 (1999)

\bibitem{Duan2001}
Duan, L.M., Lukin, M., Cirac, J.I., Zoller, P.: Long-distance quantum
  communication with atomic ensembles and linear optics.
\newblock Nature \textbf{414}(6862), 413 (2001)

\bibitem{Jiang2014}
Jiang, L., Muralidharan, S., Kim, J., Lutkenhaus, N., Lukin, M.: Ultrafast and
  \textsc{F}ault-\textsc{T}olerant \textsc{Q}uantum \textsc{C}ommunication over
  long distances.
\newblock In: Frontiers in Optics, pp. FTu1A--1. Optical Society of America
  (2014)

\bibitem{Curty2004}
Curty, M., Lewenstein, M., L{\"u}tkenhaus, N.: Entanglement as a precondition
  for secure quantum key distribution.
\newblock Phys. Rev. Lett. \textbf{92}(21), 217,903 (2004)

\bibitem{Jennewein2000}
Jennewein, T., Simon, C., Weihs, G., Weinfurter, H., Zeilinger, A.: Quantum
  cryptography with entangled photons.
\newblock Phys. Rev. Lett. \textbf{84}(20), 4729 (2000)

\bibitem{Hsieh2003}
Hsieh, M., Kempe, J., Myrgren, S., Whaley, K.B.: An explicit universal gate-set
  for exchange-only quantum computation.
\newblock Quantum Inf. Process. \textbf{2}(4), 289--307 (2003)

\bibitem{Shi2011}
Shi, J., Shi, R., Tang, Y., Lee, M.H.: A multiparty quantum proxy group
  signature scheme for the entangled-state message with quantum fourier
  transform.
\newblock Quantum Inf. Process. \textbf{10}(5), 653--670 (2011)

\bibitem{Moehring2007}
Moehring, D., Maunz, P., Olmschenk, S., Younge, K., Matsukevich, D., Duan,
  L.M., Monroe, C.: Entanglement of single-atom quantum bits at a distance.
\newblock Nature \textbf{449}(7158), 68 (2007)

\bibitem{Bernien2013}
Bernien, H., Hensen, B., Pfaff, W., Koolstra, G., Blok, M., Robledo, L.,
  Taminiau, T., Markham, M., Twitchen, D., Childress, L., et~al.: Heralded
  entanglement between solid-state qubits separated by three metres.
\newblock Nature \textbf{497}(7447), 86 (2013)

\bibitem{Hofmann2012}
Hofmann, J., Krug, M., Ortegel, N., G{\'e}rard, L., Weber, M., Rosenfeld, W.,
  Weinfurter, H.: Heralded entanglement between widely separated atoms.
\newblock Science \textbf{337}(6090), 72--75 (2012)

\bibitem{Eghbali2017}
Eghbali-Arani, M., Ameri, V.: Entanglement of two hybrid optomechanical
  cavities composed of bec atoms under bell detection.
\newblock Quantum Inf. Process. \textbf{16}(2), 47 (2017)

\bibitem{Zukowski1993}
Zukowski, M., Zeilinger, A., Horne, M.A., Ekert, A.K.: Event-ready-detectors
  \textsc{B}ell experiment via entanglement swapping.
\newblock Phys. Rev. Lett. \textbf{71}, 4287--4290 (1993)

\bibitem{Ghasemi2016}
Ghasemi, M., Tavassoly, M.K.: Entanglement swapping to a qutrit-qutrit atomic
  system in the presence of \textsc{K}err medium and detuning parameter.
\newblock Eur. Phys. J. Plus \textbf{131}(9), 297 (2016)

\bibitem{Ghasemi2017}
Ghasemi, M., Tavassoly, M.K., Nourmandipour, A.: Dissipative entanglement
  swapping in the presence of detuning and \textsc{K}err medium: \textsc{B}ell
  state measurement method.
\newblock Eur. Phys. J. Plus \textbf{132}(12), 531 (2017)

\bibitem{Nourmandipour2016}
Nourmandipour, A., Tavassoly, M.K.: Entanglement swapping between dissipative
  systems.
\newblock Phys. Rev. A \textbf{94}(2), 022,339 (2016)

\bibitem{nourmandipour2015}
Nourmandipour, A., Tavassoly, M.K.: A novel approach to entanglement dynamics
  of two two-level atoms interacting with dissipative cavities.
\newblock Eur. Phys. J. Plus \textbf{130}(7), 148 (2015)

\bibitem{Liao2011}
Liao, Q., Fang, G., Wang, Y., Ahmad, M., Liu, S.: Entanglement swapping in two
  independent \textsc{J}aynes-\textsc{C}ummings models.
\newblock Eur. Phys. J. D \textbf{61}(2), 475--479 (2011)

\bibitem{Haroche1999}
Haroche, S.: Cavity quantum electrodynamics: a review of rydberg atom-microwave
  experiments on entanglement and decoherence.
\newblock In: AIP Conference Proceedings, vol. 464, pp. 45--66. AIP (1999)

\bibitem{Chun2006}
Chun-Hua, Y., Yong-Cheng, O., Zhi-Ming, Z.: Entanglement swapping with atoms
  separated by long distance.
\newblock Chin. Phys. \textbf{15}(8), 1793 (2006)

\bibitem{Yang2005}
Yang, M., Zhao, Y., Song, W., Cao, Z.L.: Entanglement concentration for unknown
  atomic entangled states via entanglement swapping.
\newblock Phys. Rev. A \textbf{71}(4), 044,302 (2005)

\bibitem{Pakniat2016}
Pakniat, R., Tavassoly, M.K., Zandi, M.H.: A novel scheme of hybrid
  entanglement swapping and teleportation using cavity \textsc{QED} in the
  small and large detuning regimes and quasi-\textsc{B}ell state measurement
  method.
\newblock Chin. Phys. B \textbf{25}(10), 100,303 (2016)

\bibitem{Pakniat2017}
Pakniat, R., Tavassoly, M.K., Zandi, M.H.: Entanglement swapping and
  teleportation based on cavity \textsc{QED} method using the nonlinear
  atom-field interaction: \textsc{C}avities with a hybrid of coherent and
  number states.
\newblock Opt. Commun. \textbf{382}, 381--385 (2017)

\bibitem{Jaynes1963}
Jaynes, E.T., Cummings, F.W.: Comparison of quantum and semiclassical radiation
  theories with application to the beam maser.
\newblock Proc. IEEE \textbf{51}(1), 89--109 (1963)

\bibitem{Tavis1968}
Tavis, M., Cummings, F.W.: Exact solution for an
  \textsc{N}-molecule—radiation-field \textsc{H}amiltonian.
\newblock Phys. Rev. \textbf{170}(2), 379 (1968)

\bibitem{Di2008}
Di~Fidio, C., Vogel, W., Khanbekyan, M., Welsch, D.G.: Photon emission by an
  atom in a lossy cavity.
\newblock Phys. Rev. A \textbf{77}(4), 043,822 (2008)

\bibitem{Luong2016}
Luong, D., Jiang, L., Kim, J., L{\"u}tkenhaus, N.: Overcoming lossy channel
  bounds using a single quantum repeater node.
\newblock Appl. Phys. B \textbf{122}(4), 96 (2016)

\bibitem{Li2016}
Li, L., Albert, V.V., Michael, M., Muralidharan, S., Zou, C., Jiang, L.:
  Quantum repeater with continuous variable encoding.
\newblock In: APS Division of Atomic, Molecular and Optical Physics Meeting
  Abstracts (2016)

\bibitem{Wang2012}
Wang, S., Chen, W., Guo, J.F., Yin, Z.Q., Li, H.W., Zhou, Z., Guo, G.C., Han,
  Z.F.: 2 \textsc{GH}z clock quantum key distribution over 260 km of standard
  telecom fiber.
\newblock Opt. lett. \textbf{37}(6), 1008--1010 (2012)

\bibitem{Van2009}
Van~Meter, R., Ladd, T.D., Munro, W., Nemoto, K.: System design for a long-line
  quantum repeater.
\newblock IEEE/ACM Trans. Netw. \textbf{17}(3), 1002--1013 (2009)

\bibitem{Yuan2008}
Yuan, Z.S., Chen, Y.A., Zhao, B., Chen, S., Schmiedmayer, J., Pan, J.W.:
  Experimental demonstration of a bdcz quantum repeater node.
\newblock Nature \textbf{454}(7208), 1098 (2008)

\bibitem{Xu2017}
Xu, P., Yong, H.L., Chen, L.K., Liu, C., Xiang, T., Yao, X.C., Lu, H., Li,
  Z.D., Liu, N.L., Li, L., et~al.: Two-hierarchy entanglement swapping for a
  linear optical quantum repeater.
\newblock Phys. Rev. Lett. \textbf{119}(17), 170,502 (2017)

\bibitem{Chen2017}
Chen, L.K., Yong, H.L., Xu, P., Yao, X.C., Xiang, T., Li, Z.D., Liu, C., Lu,
  H., Liu, N.L., Li, L., et~al.: Experimental nested purification for a linear
  optical quantum repeater.
\newblock Nat. Photonics \textbf{11}(11), 695 (2017)

\bibitem{Fidio2009}
\textcolor{blue}{Di Fidio, C and Vogel, W}: \textcolor{blue}{Entanglement
  signature in the mode structure of a single photon}.
\newblock \textcolor{blue}{Phys. Rev. A}
  \textbf{\textcolor{blue}{79}}(\textcolor{blue}{5}), \textcolor{blue}{050,303}
  (\textcolor{blue}{2009})

\bibitem{Man2011}
\textcolor{blue}{Man, ZX and Xia, YJ and An, NB}: \textcolor{blue}{Quantum
  dissonance induced by a thermal field and its dynamics in dissipative
  systems}.
\newblock \textcolor{blue}{Eur. Phys. J. D}
  \textbf{\textcolor{blue}{64}}(\textcolor{blue}{2-3}),
  \textcolor{blue}{521--529} (\textcolor{blue}{2011})

\bibitem{Grunwald2011}
\textcolor{blue}{Gr{\"u}nwald, P and Singh, SK and Vogel, W}:
  \textcolor{blue}{Raman-assisted Rabi resonances in two-mode cavity QED}.
\newblock \textcolor{blue}{Phys. Rev. A}
  \textbf{\textcolor{blue}{83}}(\textcolor{blue}{6}), \textcolor{blue}{063,806}
  (\textcolor{blue}{2011})

\bibitem{Zhang2010}
\textcolor{blue}{Zhang, Guo-Feng and Xie, Xin-Chen}:
  \textcolor{blue}{Entanglement between two atoms in a damping Jaynes-Cummings
  model}.
\newblock \textcolor{blue}{Eur. Phys. J. D}
  \textbf{\textcolor{blue}{60}}(\textcolor{blue}{2}),
  \textcolor{blue}{423--427} (\textcolor{blue}{2010})

\bibitem{Baghshahi2016}
\textcolor{blue}{Baghshahi, H R and Tavassoly, M K and Behjat, A}:
  \textcolor{blue}{Entanglement of a damped non-degenerate $\Diamond$-type atom
  interacting nonlinearly with a single-mode cavity}.
\newblock \textcolor{blue}{Eur. Phys. J. Plus}
  \textbf{\textcolor{blue}{131}}(\textcolor{blue}{4}), \textcolor{blue}{80}
  (\textcolor{blue}{2016})

\bibitem{Shore1993}
\textcolor{blue}{Shore, Bruce W and Knight, Peter L}: \textcolor{blue}{The
  jaynes-cummings model}.
\newblock \textcolor{blue}{J. Mod. Opt.}
  \textbf{\textcolor{blue}{40}}(\textcolor{blue}{7}),
  \textcolor{blue}{1195--1238} (\textcolor{blue}{1993})

\bibitem{Barnett2007}
\textcolor{blue}{Barnett, Stephen M and Jeffers, John}: \textcolor{blue}{The
  damped Jaynes--Cummings model}.
\newblock \textcolor{blue}{J. Mod. Opt.}
  \textbf{\textcolor{blue}{54}}(\textcolor{blue}{13-15}),
  \textcolor{blue}{2033--2048} (\textcolor{blue}{2007})

\bibitem{Zheng2000}
Zheng, S.B., Guo, G.C.: Efficient scheme for two-atom entanglement and quantum
  information processing in cavity \textsc{QED}.
\newblock Phys. Rev. Lett. \textbf{85}(11), 2392 (2000)

\bibitem{Ai2007}
Ai-Xi, C., Li, D.: Entanglement swapping between atom and cavity and generation
  of entangled state of cavity fields.
\newblock Chin. Phys. \textbf{16}(4), 1027 (2007)

\bibitem{Wootters1998}
Wootters, W.K.: Entanglement of formation of an arbitrary state of two qubits.
\newblock Phys. Rev. Lett. \textbf{80}(10), 2245 (1998)

\end{thebibliography}

   \end{document}